\documentclass[aps,longbibliography,twocolumn,nofootinbib, superscriptaddress]{revtex4-2}
\usepackage{amsmath}
\usepackage{amssymb}
\usepackage{graphicx}
\usepackage{comment}
\usepackage{soul,xcolor}
\usepackage{bbold}
\usepackage[colorlinks=true ,urlcolor=blue,urlbordercolor={0 1 1}]{hyperref}
\usepackage{tikz}
\usepackage{braket}
\usepackage[english]{babel}
\usepackage{graphicx}
\usepackage{lipsum}
\usepackage{xr}
\usepackage{ulem}
\newenvironment{blueblock}
  {\begingroup\color{black}}{\endgroup}
  
\newcommand{\<}{\langle}

\newcommand{\down}{\downarrow}
\renewcommand{\>}{\rangle}
\renewcommand{\(}{\left(}
\renewcommand{\)}{\right)}
\renewcommand{\[}{\left[}
\renewcommand{\]}{\right]}
\renewcommand{\v}[1]{\mathbf{#1}} 

\renewcommand{\d}{\partial}

\newcommand{\eps}{\epsilon}

\newcommand{\be}{\begin{equation}}
\newcommand{\ba}{\begin{align}}
\newcommand{\ee}{\end{equation}}
\newcommand{\bea}{\begin{eqnarray}}
\newcommand{\eea}{\end{eqnarray}}
\newcommand{\beq}{\begin{equation}}
\newcommand{\eeq}{\end{equation}}
\newcommand{\beqn}{\begin{eqnarray}}
\newcommand{\eeqn}{\end{eqnarray}}

\newcommand{\bfk}{\mathbf{k}}
\newcommand{\bfq}{\mathbf{q}}
\newcommand{\bfr}{\mathbf{r}}

\newcommand{\moire}{moir\'e }

\newcommand{\tb}[1]{ \textcolor{black} }

\renewcommand{\d}{^{\dagger}}

\renewcommand{\L}{\mathcal{L}}

\begin{document}

\title{Unconventional superconductivity mediated by exciton density wave fluctuations}

\author{Ajesh Kumar}
\thanks{These authors contributed equally to this work.}
\affiliation{Department of Physics, Massachusetts Institute of Technology, Cambridge, Massachusetts 02139, USA}
\author{Adarsh S. Patri}
\thanks{These authors contributed equally to this work.}
\affiliation{Department of Physics, Massachusetts Institute of Technology, Cambridge, Massachusetts 02139, USA}
\affiliation{Department of Physics and Astronomy \& Stewart Blusson Quantum Matter Institute, University of British Columbia, Vancouver BC, Canada, V6T 1Z4}
\author{T. Senthil}
\affiliation{Department of Physics, Massachusetts Institute of Technology, Cambridge, Massachusetts 02139, USA}

\date{\today}

\begin{abstract}
Synthetic platforms afford an unparalleled degree of controllability in realizing strongly-correlated phases of matter. 
In this work, we study the possibility of electrically tunable exciton-mediated superconductivity arising in charge-imbalanced bilayer semiconductors. 
Focusing on the case of a bilayer semiconductor heterostructure, we identify the gating conditions required to achieve exciton density wave order within a self-consistent Hartree-Fock approximation. 
We analyze the role of the coupling of excitonic fluctuations to the fermionic charge carriers to find that the Goldstone mode of the density wave order can mediate attractive interactions leading to superconductivity. Furthermore, when the system is close to the density wave ordering, the interactions mediated by low-energy exciton modes can support an interlayer pair-density wave superconductor of anisotropic character. We discuss experimental signatures associated with these phenomena. 

\end{abstract}
\maketitle

The recent advent of artificial, synthetic platforms have provided a fertile ground for the controlled exploration of a myriad of strongly-correlated electronic phases of matter.
Remarkably, in a number of synthetic multilayer graphene and transition metal dichalcogenide (TMD) systems, fascinating phenomena -- including electron nematicity, exciton condensates, superconductivity, strange metallicity, and quantum anomalous Hall effects -- have already been discovered, in some cases in a single system (for a sample of papers see Refs. 
\cite{cao_pablo_ci_tbg, cao_pablo_sc_tbg,sharpe2019emergent, serlin2020intrinsic, wu2021chern, sc_nematicity_tbg, oh2021evidence, fci_tbg_2021, linear_t_2019, sm_tbg_pablo, sm_tbg_efetov, 
feng_wang_abc_tlg, Nuckolls_2023, lu2023fractionalquantumanomaloushall, han2024signatureschiralsuperconductivityrhombohedral,tang2019wse2ws2moiresuperlatticesnew, Regan_2020, Wang_2020, Li_2021_tmd, Huang_2021, Li_2021, Ghiotto_2021, Zhao_2023, cai2023signatures, park2023observation, tao_vc_qah_tmd, kang2024observationfractionalquantumspin,liu2024opticalsignaturesinterlayerelectron}); for overarching reviews on graphene and TMD based \moire materials see Refs. \cite{manzeli20172d, andrei2021marvels,Nuckolls_2024} ).
These systems afford a level of tunability and control that is arguably unparalleled, especially when compared to archetypal strong-correlated systems where one is largely restricted to chemical doping or pressure tuning for the more stoichiometrically rigid systems.
Synthetic platforms, on the other hand, possess additional knobs, such as an external electric (displacement) field or an intra-layer dependent bias potential, that can be applied, which open a new avenue into the types of phases that can be realized as well as possible phase transitions \cite{song_phasetransitions_2024}.
Compounded with the variety of heterostructure combinations, a rich variety of correlated phases can be realized and studied in a deliberate fashion, which may help to shine light on the outstanding questions in the field of strongly-correlated electronic systems.

Motivated to examine synthetic platforms that realize proximate phases and phase transitions, we focus on spatially separated doped bilayer semiconductors with strong electrically tunable interlayer correlations.
This metallic setup is conducive to interlayer exciton condensation.
Indeed, such a setting is particularly fertile as (i) the exciton condensate phase can be uniform in real-space or have additional translational symmetry breaking (with a finite-momentum ordering order parameter), and (ii) can potentially host a continuous exciton condensation phase transition.
This finite-momentum ordering is reminiscent of the finite-$\v{Q}$ spin-density wave phase transition, which has a storied and admittedly complicated abundance of studies (see Refs.~\cite{lohneysen2007,sslee_review_nfl_2018} and references therein).
In our study, we focus on \textcolor{black}{generic} type-II band aligned bilayer semiconductors separated by an insulating barrier \textcolor{black}{(such as a  WS$_2$/WTe$_2$ heterostructure)}, that are dual-gated and separately contacted, allowing for the application of an interlayer bias voltage $V_b$. 
Such a system enables conduction electrons and holes to reside natively in their own separated layers, thus forming a $p-n$ junction. 
Here, type-II indicates a `staggered' alignment of the conduction and valence bands from each layer~\cite{lo2011emergent, ozcelik_2016_band_alignment, sun2021recent, kistner_morris_2024} that is favorable to the formation of interlayer excitons.


\begin{figure}[t]
\includegraphics[width = 0.4\textwidth]{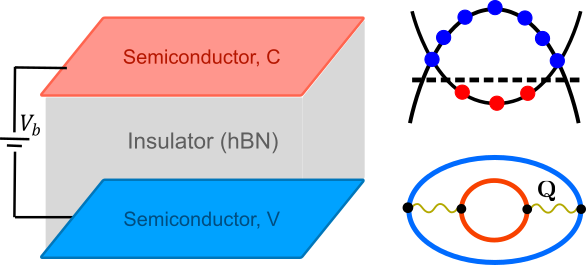}
\caption{Left: Schematic of semiconductor $-$ hBN spacer $-$ semiconductor sandwich.
The densities on the individual layer are controlled independently with an external bias potential $V_b$ applied staggerdly on each layer. 
Right top: Schematic of the unequally occupied conduction and valence bands in the electron-hole plasma state.
Right bottom: Schematic of the Fermi surfaces, where the red (blue) is an electron (hole)-like Fermi surface. The exciton density wave order is triggered by the condensation of electron-hole pairs connected by the wave-vector $|\v Q|$.} 
\label{fig_schematic}
\end{figure}

\textit{Microscopic model of $p-n$ junction.---} We consider a gate-tunable $p-n$ junction structure, wherein two semiconducting monolayers are separated by an insulating barrier, as shown in Fig. \ref{fig_schematic}.
The top and bottom layers are biased externally 
to overcome the native bandgap of the individual semiconductors.
We will hereafter measure $V_b$ relative to the value required to achieve band touching.
Moreover, the monolayers are connected to their own electronic reservoir~\cite{zheng1997,zeng2020,tu2024}, 
which enables the electronic density on each layer to be individually tuned.
Specifically, we examine the situation where the top/bottom layers are electron/hole doped.
The corresponding kinetic energy 
is,
$H^0_{c/v} = \sum_{\v k} \xi_{c/v, \v k}  f\d_{c/v,\v k} f_{c/v,\v k}$ ,
where the conduction and valence band dispersions
are $C_{2z}$ symmetric and
given by $\xi_{c,\v k} = \left(\frac{k_x^2}{2m_{c,x}} + \frac{k_y^2}{2m_{c,y}} - \frac{V_b}{2} -\mu \right)$ and $\xi_{v,\v k} = -\left(\frac{k_x^2}{2m_{v,x}} + \frac{k_y^2}{2m_{v,y}} - \frac{V_b}{2}  + \mu \right)$, where $m_{c/v, x/y}$ are the anisotropic effective masses of the electrons in the conduction (top) and valence (bottom) layers, and $\mu$ is the chemical potential. 

Within each layer there is in-principle a two-fold spin degeneracy (or alternatively valley degeneracy as the valley is locked to spin via spin-orbit coupling \cite{manzeli20172d}), which introduces additional possibilities for the spin-correlations of the exciton condensate. 
We circumvent this complication, by drawing motivation from a recent experimental study \cite{nguyen2023degeneratetrionliquidatomic}, where spin-polarized excitons have been reported in Coulomb-coupled monolayers of MoSe$_2$ and WSe$_2$; indeed mean-field studies of charge-compensated ${\v Q = 0}$ exciton condensates also indicate that spin-polarization is stabilized by an infinitesimal Zeeman field~\cite{Wu_2015}.
As such, we focus on an analogous situation, where a spin-polarized exciton condensate forms (for example, spin-$\uparrow$ from the conduction/top layer and spin-$\down$ from the valence/bottom layer).
In this sense, the spin-layer labels are `locked'.
With this understanding, we henceforth drop the spin-labels and keep the layer labels in what follows.
\textcolor{black}{Although the spin-polarized exciton condensate breaks the underlying $C_{2z}$ symmetry, a combined operation ($\equiv C'_{2z}$) of $C_{2z}$ \textit{and} a $\pi$ spin-rotation about $\hat{x}$ remains a symmetry.}


The fermions in each layer interact via a repulsive dual gate-screened Coulomb interaction, 
\begin{align}
\label{eq_coulomb_repulsion}
    H_{int} = \frac{1}{2A} \sum_{\v q}  V(\v q) \Big[ & n_c(\v q) n_c(-\v q) + n_v(\v q) n_v(-\v q)  \Big. \nonumber \\
    & + \Big. 2e^{-qd} n_c(\v q) n_v(-\v q)  \Big], 
\end{align}
where $A$ is the area of the system, 
$V(\v q) = \frac{e^2}{2\epsilon_0 \epsilon q} \text{tanh}(q d_{g})$
is the screened Coulomb interaction for a dual-gated setup with an inter-gate distance $d_g$, and interlayer separation $d$.
We have compactly defined the conduction and valence densities, $n_{c/v}(\v q) = \sum_{\v k}f\d_{c/v,\v k+\v q} f_{c/v,\v k}$.
The system has a combined $U(1)_c \times U(1)_v$ symmetry, respectively corresponding to the independent conservation of particle number in each layer.
The low-energy continuum model also possesses continuous translational symmetry, which reduces to discrete translational symmetries once a lattice potential is introduced.
The interaction terms on the first line of Eq. \ref{eq_coulomb_repulsion} describe the intra-layer interactions, while the last term denotes the inter-layer interactions.
Recent experiments \cite{wang2019evidence,ma2021strongly,shi2022bilayer,nguyen2023degeneratetrionliquidatomic,perfect_coulomb_drag_shan_mak} appear to realize inter-layer excitonic instabilties, which suggests that 
intra-layer interactions play the benign role of renormalizing the dispersion (and effective masses) and the native bandgap of the semiconductors; as such its role in triggering additional (within each layer) electronic instabilities is disregarded.

\label{sec_ex_dw}

\begin{figure}[t]
\includegraphics[width = 0.22\textwidth]{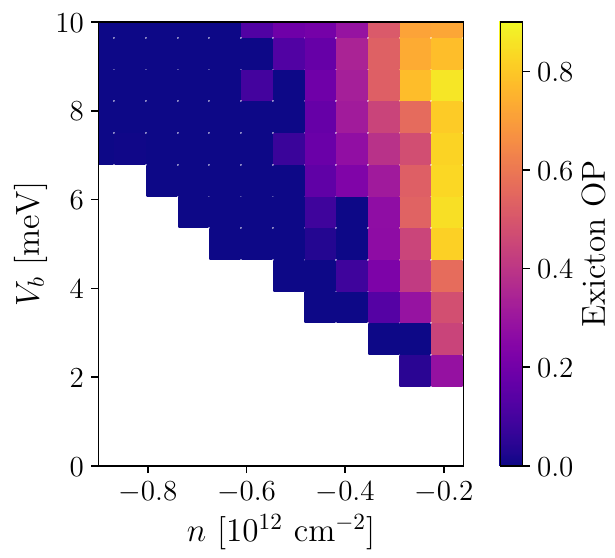}
\includegraphics[width = 0.22\textwidth]{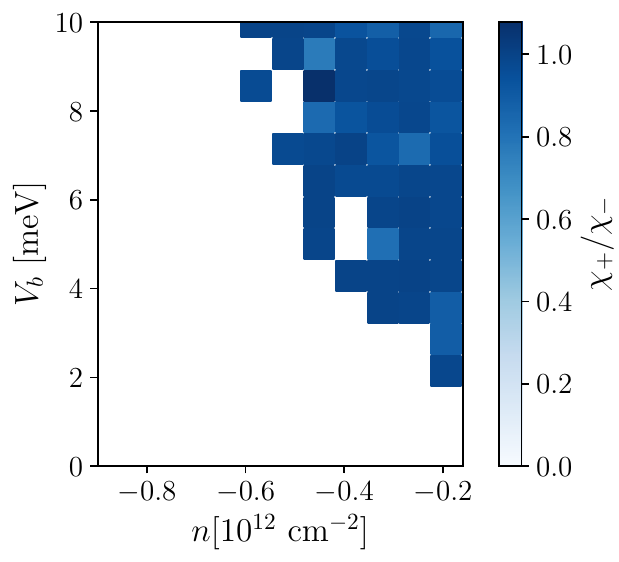} \\
\caption{Self-consistent Hartree-Fock exciton order parameter at total density $n$ and bias voltage $V_b$. 
Left: Inversion symmetric exciton order parameter $\chi = (\chi_+ + \chi_-)$, where $\chi_{\pm}$ is the exciton order parameter for $\pm \textbf{Q}$ momentum.
Right: Ratio of $+\textbf{Q}$ and $-\textbf{Q}$ exciton ordering.
Momentum mesh of 15$\times$15, and up-to 7$\times$2 (including both conduction and valence) folded bands was used.}
\label{fig_hf_main}
\end{figure}

\textit{Exciton density wave order.---}The non-intersecting electron and hole Fermi surfaces in the respective top and bottom layers are susceptible to an exciton-formation instability
of a finite-momentum (incommensurate) exciton condensation, as has been noted in previous works~\cite{pieri2007effects,varley2016}; the exciton condensed phase will be referred to as an exciton density wave (X-DW)~\cite{bi2021excitonic}.
In particular, with an unequal number of electron and hole carriers, the respective circular (elliptical) electron(hole) Fermi surfaces are separated by finite-momentum, which leads to the formation of finite-momentum excitons with a preferred center-of-mass momentum, $\v Q$, as depicted in Fig. \ref{fig_schematic}. 
The directional dependence of the center-of-mass momentum is explicitly confirmed \cite{supplemental_material}, 
within a random phase approximation (RPA).
The RPA susceptibility is peaked when the exciton center-of-mass momentum is $\pm \v Q$, where $\v Q = (k_{F,v,x}-k_{F,c,x}) \hat{x}$ where $k_{F,v/c,x}$ are Fermi momenta in the two layers along the $x$-direction.
\textcolor{black}{With continuous rotational symmetry, exciton modes soften on a circle of points in momentum; the mass anisotropy lifts this degeneracy, leading to softening at discrete momenta $\pm \v Q$. }

\begin{figure}[t]
\includegraphics[width = \columnwidth]{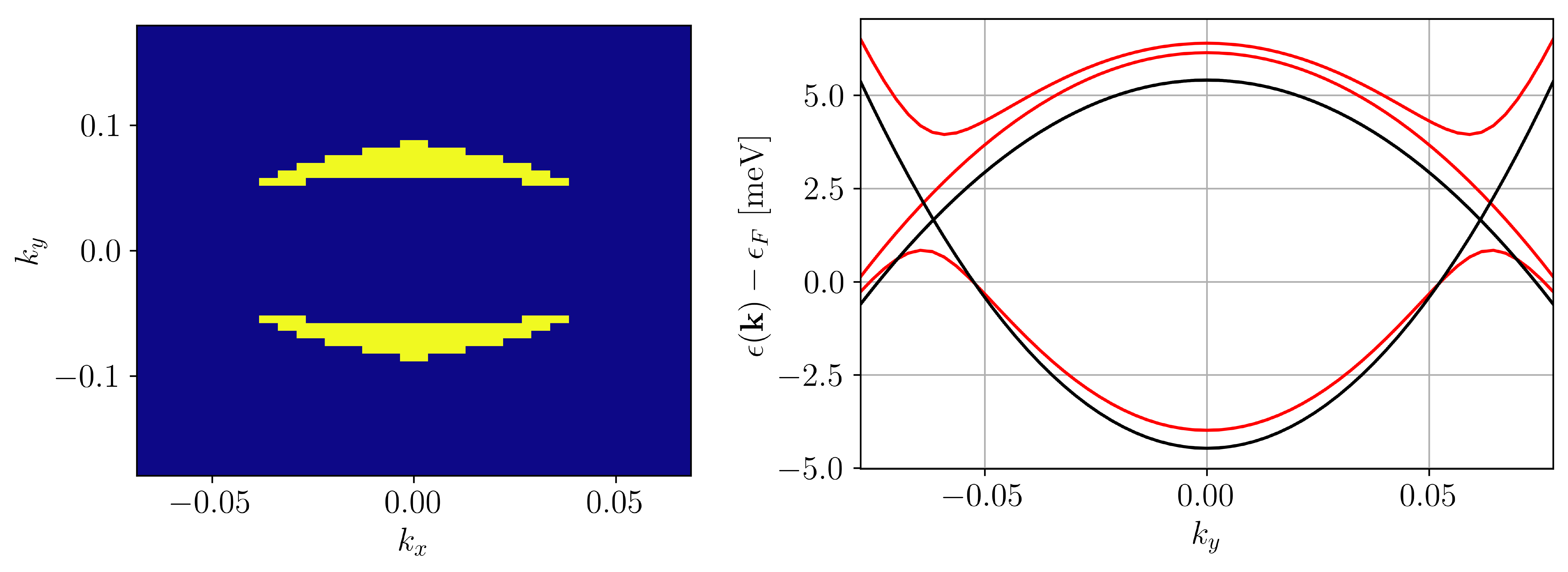}
\caption{Development of collinear exciton density wave for $n=-0.7\times 10^{12}$ cm$^{-2}$ and $V_b = 12$ meV within a three-band Hubbard approximation (valence band for ${\v k \pm \v Q}$ and conduction band for ${\v k}$).
Left: Electronic Fermi surface Fermi surface in the collinear X-DW phase (yellow denotes the filled Fermi sea). Right: Electronic Bandstructure for $k_x = 0$ in the presence/absence of the collinear X-DW in red/black.
}
\label{fig_xdw}
\end{figure}

Figure~\ref{fig_hf_main}(a) depicts the Hartree-Fock phase diagram, which indicates the onset of exciton finite-momentum exciton instabilities, $\< f\d_{v,\v k\pm\v Q} f_{c,\v k} \>$.
In real-space, the exciton density wave 
order parameter is given by $\chi_{\v r}(\v R)=\chi_+(\v R) e^{i\v Q \cdot \v r} + \chi_-(\v R) e^{-i\v Q \cdot \v r}$, where $\v r$ is a center-of-mass coordinate and $\v R$ is a relative coordinate between the electron and hole.
As shown in Fig. \ref{fig_hf_main}(b), we find $\chi_+ \approx \chi_- \equiv \chi_0$, so that the strength of the order parameter acquires a periodic spatial modulation in $\v r$, with the phase staying uniform, resulting in the spontaneous breaking of $U(1)_r$ and continuous translational symmetry along the $x$-direction. 
We refer to this as the collinear X-DW. 
We note that the density range over which this collinear X-DW exists is for $|n| \lesssim 0.6 \times 10^{12}$ cm$^{-2}$ as shown in Fig. \ref{fig_hf_main}.
We present in Fig. \ref{fig_xdw} a representative electronic bandstructure and Fermi surface depicting the impact of the exciton condensate, within a three-band Hubbard approximation (see Fig.~A.1 in \cite{supplemental_material} for the self-consistent Hartree-Fock calculation that incorporates dual-gated Coulomb interactions).
Importantly, despite a gap opening at certain momentum points in the Brillouin zone, a Fermi surface still remains.

\textit{Goldstone fluctuations mediated nodal superconductivity.---}In the collinear X-DW condensate, 
breaking of the two continuous symmetries leads to two independent Goldstone modes. 
associated with: a uniform $U(1)_r$ phase rotation, $\chi = \chi_0 \cos (\v Q \cdot \v r) \rightarrow e^{i\theta}\chi_0 \cos (\v Q \cdot \v r)$, and a uniform translation $\chi = \chi_0 \cos (\v Q \cdot \v r) \rightarrow \chi_0 \cos (\v Q \cdot \v r + \alpha)$.
While the coupling of these Goldstone modes to the fermions retains the integrity of the Fermi liquid quasiparticles~\cite{Watanabe_2014}, we ask here whether they can mediate superconductivity.
A similar theoretical study was performed in the context of inter-valley coherent fluctuation-mediated superconductivity in twisted bilayer graphene \cite{po_origin_2018, kozii_ivc_sc_tbg_2022} and multilayer graphene \cite{chatterjee2022inter, vituri2024incommensurateintervalleycoherentstates}.
For related works, see also Refs.~\cite{blinov2022interlayer,jiang2023possible}; \textcolor{black}{as well, Refs.~\cite{zerba2024realizing,von2023superconductivity} for analyses of superconductivity arising from exciton fluctuations.}

\begin{figure}[t]
\includegraphics[width = 0.7\columnwidth]{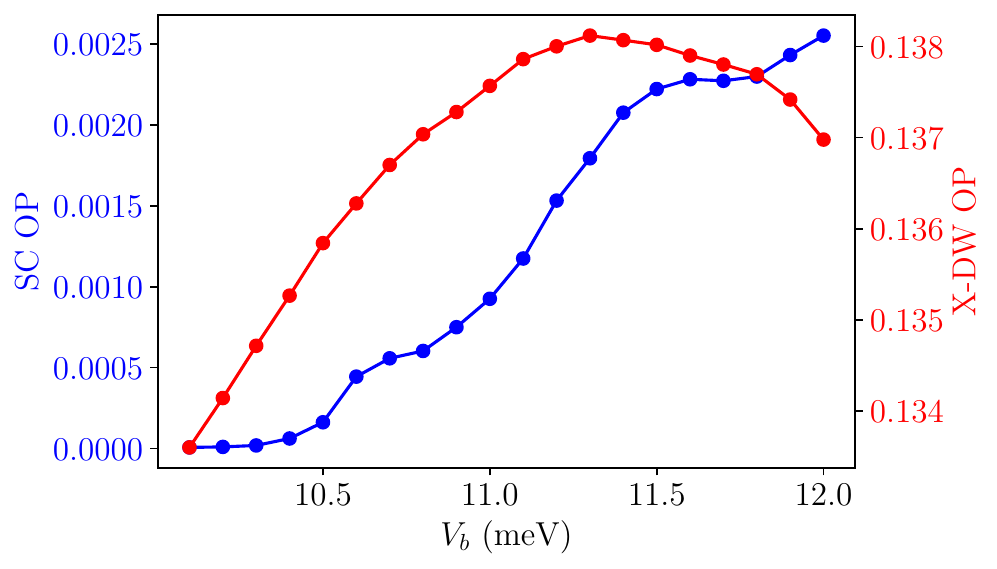}
\includegraphics[width = 0.7\columnwidth]{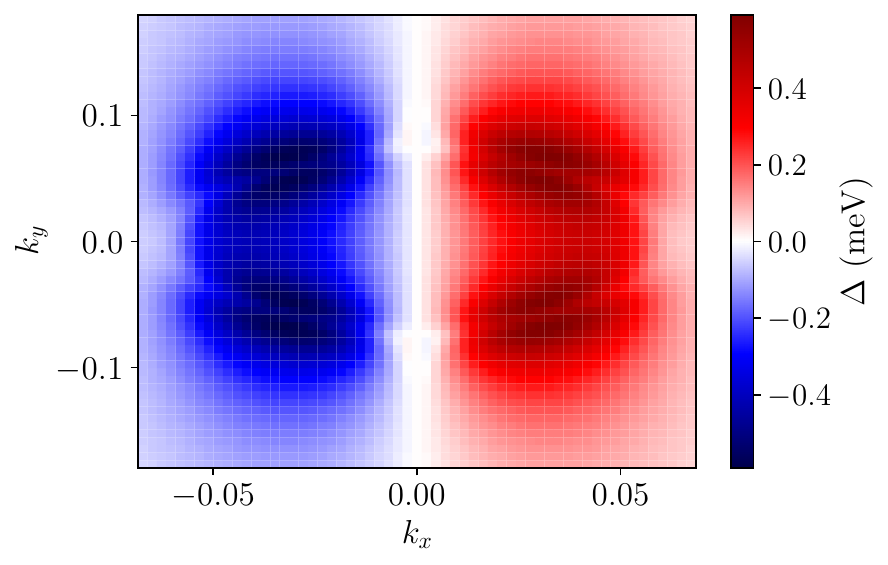}
\caption{Goldstone mode ($\delta \chi_p$)-mediated superconductivity inside the collinear X-DW phase. Top: Superconducting order parameter as a function of $V_b$ for $n=-0.7\times 10^{12}$ cm$^{-2}$. Bottom: BCS gap function as a function of momentum for $V_b=12$ meV.
Momentum mesh of $59 \times 59$, $\eps = 10$, and momentum space cutoff along the $x$-direction $k_{c,x} = 0.2$.}
\label{fig_goldstone}
\end{figure}

We confine ourselves to the consideration of static fluctuations of the collinear X-DW order parameter: $\chi(\v r) = \chi_0 \cos({\v Q \cdot \v r}) + \delta \chi_-(\v r) e^{-i\v Q \cdot \v r}+ \delta \chi_+(\v r)e^{i \v Q \cdot \v r}$, where the first term is the mean-field (saddle-point) solution.
Retaining the important phase fluctuations ($\text{Im}[\delta \chi_{\pm} (\v r)] \neq 0$) at low energies, 
we obtain a Gaussian action for these fluctuations following the standard recipe of integrating out high-energy fermions (see \cite{supplemental_material} for details). 
Diagonalizing the $2 \times 2$ fluctuation Hamiltonian, we obtain two gapless Goldstone modes: 
(i) `in-phase' fluctuation mode ($\delta \chi_+ = \delta \chi_-$, labelled by $ \delta \chi_s$), 
which can be interpreted as a superfluid mode due to the breaking of $U(1)_r$, and (ii)  an `out-of-phase' mode
 ($\delta \chi_+ = -\delta \chi_-$, labelled by $\delta \chi_{p}$), which can be interpreted as an acoustic phonon due to breaking of the continuous-translation symmetry by the fluctuations.
The corresponding Goldstone mode Hamiltonian is, 
\begin{align}
H_{G} =  \frac{\rho_{G}}{2} \sum_{\v q} (q_x^2 + \kappa_G q_y^2)\delta \chi_G(-\v q) \delta \chi_G(\v q),
\label{eq:goldstone}
\end{align}
where $G=s/p$ labels the two (normal) Goldstone modes, $\rho_G$ is the Goldstone mode stiffness and $\kappa_G$ is the anisotropy parameter, both of which are numerically calculated. 
We note that we have disregarded the dynamics of the Goldstone mode here, which when treated provides a more accurate estimates of the critical temperature (within the framework of Eliashberg theory). 
Since our focus here is on the qualitative behaviors of the superconductivity, we focus solely on the static components; similar studies of fluctuation mediated superconductivity also demonstrate the predictive power of examining the purely static component of the Goldstone action \cite{po_origin_2018, kozii_ivc_sc_tbg_2022,vituri2024incommensurateintervalleycoherentstates}.
The interactions induced by the Goldstone modes between the low-energy fermions $f$ (formed from the mean-field band)
\begin{blueblock}
in the BCS channel is given by,
\begin{align} 
\sum_G
    \frac{1}{2A\rho_{G}}\sum_{\v k, \v l} V_{BCS}^G (\v k, \v l) \ f\d_{\v k} f\d_{ -\v k} f_{-\v l} f_{\v l} .
\label{eq:goldstone_interaction}
\end{align}
We provide details in the End Matter of how the Goldstone mode low-energy fermion coupling generates this attraction.
Crucially, the superfluid mode's interaction potential is even-parity under $C'_{2z}$ ($V_{BCS}^S (\v k, \v l) = V_{BCS}^S (\v k, -\v l)$) and repulsive while the phonon mode has an odd-parity component ($V_{BCS}^P (\v k, \v l) \neq V_{BCS}^P (\v k, -\v l)$) and is attractive.
As such, the Fermi-Dirac statistics enforced odd-parity superconducting parameter (of single-flavor fermions) can only be realized via the phonon mode.
To analyze this possibility in detail, we self-consistently solve for the superconducting gap function using the interactions mediated by the phonon mode. 
Figure~\ref{fig_goldstone} depicts the superconducting order parameter (top) and the BCS gap function $\Delta(\v k)$ (bottom) of a representative parameter point obtained.
The gap function has an anisotropic $p$-wave character, with nodal points in the BdG spectrum, and inherits the time-reversal symmetry breaking of the normal X-DW state. 
While time-reversal symmetry broken superconductors are uncommon in nature, there is recent evidence for this phenomenon in rhombohedrally-stacked tetra-layer graphene~\cite{han2024signatureschiralsuperconductivityrhombohedral}.
\end{blueblock}

\textit{Critical fluctuations mediated pair-density wave.---}The (mean-field suggested continuous) critical line demarcating  the metallic electron-hole plasma and the X-DW in Fig.~\ref{fig_hf_main} indicates the development of further electronic instabilities.
In the standard lore of metallic quantum critical points, superconductivity and non-Fermi liquids are two of the competing instabilities \cite{lohneysen2007,Sachdev_2011}.
Since our goal is not to study this competition in detail~\cite{Metlitski_2015}, we focus here on the possibility of unconventional superconductivity arising from pairing of the fermions on approach from the Fermi liquid.
In \cite{supplemental_material}, we consider non-Fermi liquid behavior from the classic one-loop hot-spot model treatment which leads to a dynamical critical exponent $z=3$, and singular corrections of the electronic self-energy, $\Sigma(\omega) \sim i|\omega|^{2/3}$. 

The fluctuations about the quantum critical point are modelled by two bosonic modes $\delta \chi_{\v r} = b_{\pm} (\v r)$, which condense at the translation-symmetry breaking exciton condensate momenta $\pm \v Q$.
From the $C'_{2z}$ symmetry (that relates the $\pm\v Q$ condensates), the interaction of the bosons with the fermionic (electron/hole) excitations is captured by the Hamiltonian, $H = H_f + H_b + H_{f-b}$, where
\begin{align}
    & H_b = \sum_{s=\pm}\sum_{\v q}  \Big[c_b^2 \left((q_x - s Q)^2 + \kappa q_y^2\right) + m_b^2 \Big] |b_s(\v q)|^2 \nonumber\\
    & H_{f-b} = g\sum_{s=\pm} \sum_{\v q}   b_s(\v q) \rho_{DW}\d(\v q) + \text{h.c.}
\end{align}
Here $H_f$ is the bare action of the fermions that form electron and hole Fermi surfaces.
The density wave operator is $\rho_{DW}(\v q) = \int_{\v k} f\d_{v,\v k+\v q} f_{c,\v k}$, the mass (speed) of the bosons is given by $m_b$ ($c_b = 1$, for simplicity), $\kappa$ captures the anisotropy in the boson dispersion, and we have only kept the leading momentum independent coupling constant $g$. 
Our approximation of considering only static fluctuations is justified in the presence of a sufficiently large $m_b$.
Integrating out the bosons we arrive at an effective interaction term between the fermions, $-\frac{1}{A}\sum_{\v q} V_{DW}(\v q) \rho_{DW}(\v q, \Omega=0) \rho\d_{DW}(\v q, \Omega=0)$, where 
\begin{align}
    V_{DW}(\v q) = g^2 \sum_{s=\pm} \frac{1}{(q_x - sQ)^2 + \kappa^2 q_y^2 + m_b^2}, 
    \label{eq:sc}
\end{align}
\textcolor{black}{where the constants $g$, $\kappa$ and $m_b$ are estimated (as functions of $n$ and $V_b$) by using a Hubbard interaction that approximately reproduces the quantum critical line in Fig.~\ref{fig_hf_main} (see \cite{supplemental_material}). }

\begin{figure}[t]
\includegraphics[width = 0.7\columnwidth]{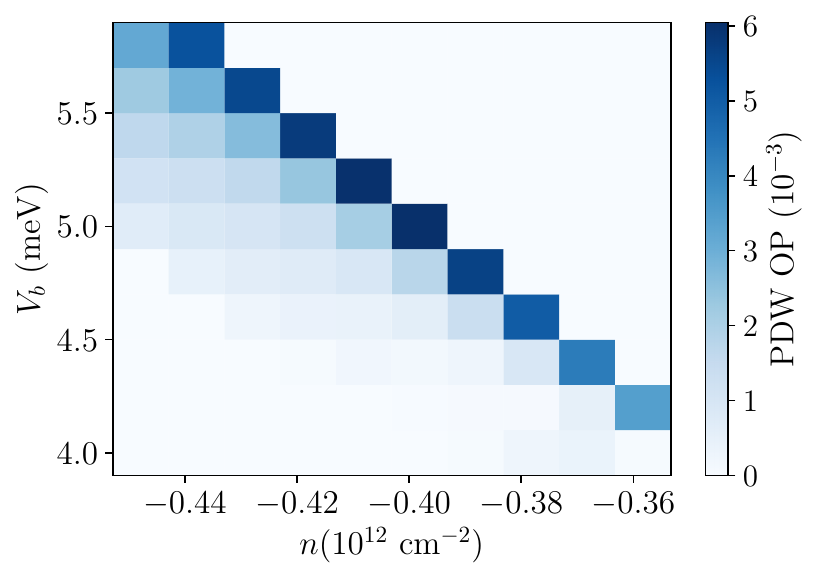}
\includegraphics[width = 0.7\columnwidth]{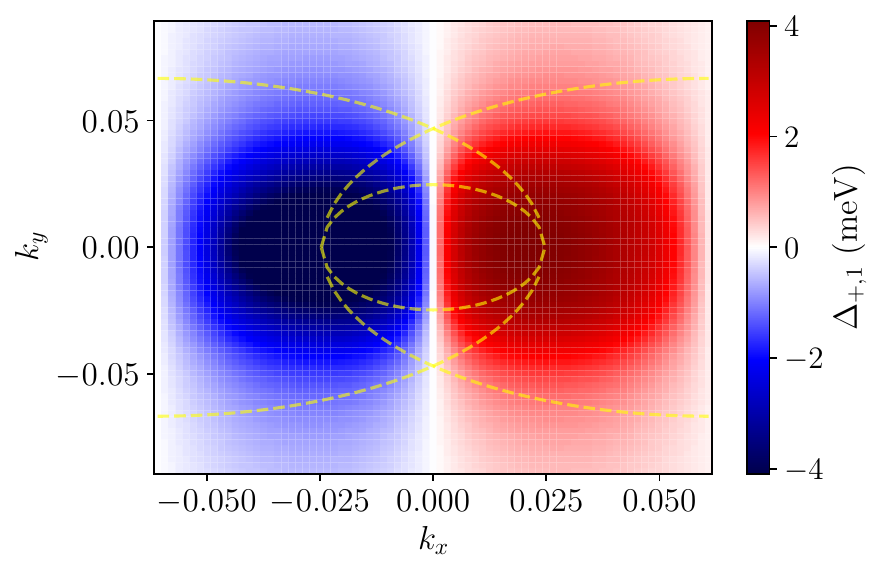} \\
\caption{Critical fluctuation-mediated pair-density wave superconductivity. Top: Order parameter of the pair-density wave (PDW) superconductor plotted on the electron-hole plasma side of the phase diagram (lower triangle). Bottom: The PDW gap function for pairing momentum $\v Q$ plotted in the first Brillouin zone. Parameters: $n=-0.36\times 10^{12}$ cm$^{-2}$ and $V_b = 4.2$ meV. Dashed lines are Fermi surfaces of $f_{c,\v k}$, $f_{v,\v k + \v Q}$ and $f_{c,\v k - \v Q}$.
Momentum mesh of $79 \times 79$. 
}
\label{fig_sc_main}
\end{figure}

The non-vanishing \textcolor{black}{inter-layer} superconducting instability is found to be a pair-density wave (PDW) state with pairing momentum $\v Q$: $\< f_{v,\v k + \v Q} f_{c,-\v k} \>, \< f_{c,\v k + \v Q} f_{v,-\v k} \> \neq 0$ (as depicted in Fig. \ref{fig_sc_main}), as opposed to the standard zero-momentum pairing.
As seen, the PDW superconductor is stabilized near the quantum critical line, where the interactions mediated by the soft bosons $b_{\pm}$ are strongest.
The development of the PDW state requires a critical interaction strength, therefore, for small values of $g$ or large values of $m_b$, superconductivity disappears, as seen in Fig.~\ref{fig_sc_main}. 
We find that the obtained PDW gap functions have an anisotropic $p$-wave character (see Fig.~\ref{fig_sc_main}, bottom panel) with nodes at $k_x=0, -Q$.
The resulting BdG spectrum consists of gapless quasiparticles that form a Fermi surface, as is typical for PDW superconductors~\cite{Agterberg_2020}.

\textit{Experimental signatures.---}
The X-DW order as well as both the superconducting phases involve an inter-layer condensate (the Goldstone mediated superconductor also involves intra-layer pairing).
This suggests drag transport -- with separate electrical contacts, current is driven in one layer and the voltage generated is measured in the opposite layer -- as a natural probe for these phenomena~\cite{eisenstein2014exciton}.
Analogous to the exciton condensates found in bilayer quantum Hall systems, the drag (and counterflow) resistivity will be zero in the X-DW phase even though the excitons condense at a finite momentum. 
We direct the reader to  \cite{supplemental_material} for arguments based on effective response actions.
Since the PDW superconductor is an interlayer Cooper pair condensate, the drag resistivity is zero even in this case. 
However, in contrast to the X-DW, the PDW state will exhibit superconductivity when the layers are contacted together.
Finally, in the quantum critical NFL regime, the hotspot contribution to AC counterflow transport could potentially contain signatures of the quantum criticality, if they can be isolated from the contributions from the other ``cold" parts of the Fermi surface~\cite{shi2024excitonicquantumcriticalitybilayer}. 

\textcolor{black}{For the predicted nodal superconductors, a direct signature is the characteristic `V-shape' of the density of states from scanning tunneling microscopy (STM)~\cite{oh2021evidence,kimh2022ttg}.
With the dual gate setup obstructing easy access the sample surface, a single-gate setup (that simultaneously tunes both density and bias potential) with an exposed top layer would allow a portion of the superconducting phase diagram to be directly imaged.
In addition, measurements of the superfluid stiffness -- via recently developed kinetic inductance techniques (see Refs. \cite{banerjee2025superfluid, tanaka2024superfluidstiffnessflatbandsuperconductivity}) -- would also indicate the gapless BdG spectrum from its linear-in-temperature scaling behavior (in the clean limit \cite{Hirschfeldpenetrationdepth,patri2025familymultilayergraphenesuperconductors}).
}



\textit{Outlook.---}
The two forms of superconductivity -- one arising due to Goldstone fluctuations and another due to quantum critical fluctuations --  can be realized at zero temperature.
A natural direction to explore would be finite temperature energy competition between the two superconductors (for instance, the critical temperature $T_c$ of each and the dependence on $n, V_b$).
\textcolor{black}{It would also be interesting to incorporate dynamical screening effects to examine the magnitude of suppression of the exciton onset temperature, as compared to the zero-momentum exciton condensate \cite{PhysRevLett.133.056501}.}

Recent experiments have presented evidence for the existence of a Fermi liquid state of trions (bound state of a fermion and an exciton)~\cite{nguyen2023degeneratetrionliquidatomic}. 
It would be interesting to incorporate trions in our framework (\textcolor{black}{expected to become important for hole-electron densities of the ratio $p=2n$ \cite{nguyen2023degeneratetrionliquidatomic}}), which would provide an alternate route for the excitons to become gapped by starting from the exciton condensed phase (as studied in the cold-atom context in Ref.~\cite{altman2012}). 

Finally, it would be intriguing to examine whether analogous unconventional superconductivity may be triggered in compensated twisted double bilayer graphene, where finite-momentum exciton condensates have been discussed~\cite{ghosh2022evidencecompensatedsemimetalelectronic,ghorai_2023}.

\section*{Acknowledgements}
We thank P. A. Lee and Y. Zeng for insightful discussions.
The authors acknowledge the MIT SuperCloud and Lincoln Laboratory Supercomputing Center for providing HPC resources that have contributed to the research results reported within this manuscript.
A.K. was supported by the Gordon and Betty Moore Foundation EPiQS Initiative through Grant No. GBMF8684 at the Massachusetts Institute of Technology.
A.S.P. was supported by a Simons Investigator Award to T.S. from the Simons Foundation at the Massachusetts Institute of Technology, and is supported by NSERC, CIFAR and by the Gordon and Betty Moore Foundation’s EPiQS Initiative through Grant No. GBMF11071 at the University of British Columbia. T.S. was supported by the Department of Energy under grant DE-SC0008739.  and partially by the
Simons Collaboration on Ultra-Quantum Matter, which is a grant from the Simons Foundation (651446, TS).

\nocite{lv2015perfect,conti2020transition,dong2023theory,berk_schrieffer_1966,Scalapino1986,senthil2014massenhancementnearoptimal}
\bibliographystyle{apsrev4-2}
\bibliography{ref_exciton}

\break
\section*{End Matter}

\begin{blueblock}
\textit{Role of superfluid and phonon modes in generating superconductivity.---}The effective interaction between the low-energy fermions $f$  (formed from the mean-field band) is generated by first rewriting the bare fluctuations ($\delta \chi_{\pm}$) in terms of the defined Goldstone modes ($\delta \chi_G$)
\begin{align}
    H_{f-G} = -i \int_{\v q, \v k}\alpha_G(\v k + \v q, \v k) \delta \chi_G(-\v q) f\d_{\v k + \v q} f_{\v k}.
\end{align}
This satisfies the Hermiticity condition: $\alpha_G^*(\v k + \v q, \v k) = -\alpha_G(\v k, \v k + \v q)$.
We next integrate out the Goldstone modes (using the quadratic action Eq. \ref{eq:goldstone} in the main text) to obtain an effective interaction term between the low-energy fermions, which we write in the BCS channel since our primary interest is in possible superconductivity:
\begin{align} 
\sum_G
    \frac{1}{2A\rho_{G}}\sum_{\v k, \v l} f\d_{\v k} f\d_{ -\v k} f_{-\v l} f_{\v l} \frac{\alpha_G(\v k,\v l)\alpha_G(-\v k, -\v l)}{(k_x-l_x)^2+\kappa_G (k_y-l_y)^2}.
\label{eq:goldstone_interaction}
\end{align}
Since we consider $\chi_0$ to be real, the mean-field wavefunctions can be chosen to be real, implying that $\alpha_G$ is real for all momenta.
The fermion-Goldstone mode coupling $\alpha_G$ satisfies: $\alpha_{s/p}(\v k, \v l) = \pm \alpha_{s/p}(-\v k, -\v l)$, since the X-DW state satisfies the $C'_{2z}$ symmetry (see Supplemental Material~\cite{supplemental_material}).
From the above interaction term, which we define as $V^G_{BCS}(\v k, \v l)$, it is readily seen that the interactions mediated by $\delta \chi_{s/p}$ is repulsive/attractive in character.
Since we are examining supeconductivity emerging from a single band of fermions, Fermi-Dirac statistics enforces an odd-parity order parameter.
However, for the superfluid mode, we find $\alpha_s(\v k, \v l) = \alpha_s(\v k, -\v l)$ implying an even-parity $V^s_{BCS}(\v k, \v l) = V^s_{BCS}(\v k, -\v l)$. Therefore, superconductivity cannot occur in this case. 

For the phonon mode, on the other hand, we find that generically $\alpha_p(\v k, \v l) \neq \alpha_p(\v k, -\v l)$. Therefore, $V^p_{BCS}$ has an odd-parity component 
and can lead to superconductivity within the single band. 
\end{blueblock}


\onecolumngrid
\newpage
\renewcommand\thefigure{A.\arabic{figure}}
\renewcommand\theequation{A.\arabic{equation}}
\setcounter{figure}{0}

\section*{Supplemental Material for ``Unconventional superconductivity mediated by exciton density wave fluctuations''}

\section{Microscopic Parameters of Model}
\label{app_params}

In this study, we take WS$_2$/WTe$_2$ as a representative material in our numerics. 
This heterostructure is expected to have a type-II band alignment with the conduction (valence) band belonging to the WS$_2$ (WTe$_2$) layer.
Let us define a microscopic lattice length scale $a = 0.32 \text{nm}$. 
This is not necessarily the lattice constant of the semiconductors under consideration, but a convenient length scale that we use to express other physical parameters.
The effective valence and conduction masses are $m_{v,x} a^2 = 0.89 \text{eV}^{-1}$, $m_{v,y} a^2 = 0.54 \text{eV}^{-1}$\cite{lv2015perfect}, $m_{c,x} a^2 = m_{c,y} a^2 = 0.33 \text{eV}^{-1}$~\cite{conti2020transition}, with a gate distance of $d_g = 1.5 \text{nm}$ and interlayer distance $d=3.5$ \text{nm}. 
We note that our parameter choice of $d_g$ is motivated by the MoSe$2$/WSe$_2$ experimental study Ref.~\cite{nguyen2023degeneratetrionliquidatomic,perfect_coulomb_drag_shan_mak}, where the W and Mo layers are separated by a thin hBN spacer of thickness of $1.5-2$nm in the channel region, while the exciton `contact' region is separated by $10-20$ nm.
For the Hubbard approximation models, we adopt a rectangular momentum grid of cutoff length $k_{c,x} = 0.1 a^{-1}$ along the $x$-direction, and $k_{c,y} = k_{c,x}\(\frac{m_{v,y}m_{c,y}}{m_{v,x}m_{c,x}}\)^{1/4}$ along the $y$-direction. 
The chemical potential and bias voltage 
defined in the main text determine the Fermi momenta $k_{F,c/v,x/y}$ of the carriers in the conduction/valence ($c$/$v$) layers along the $\hat{x}/\hat{y}$ directions.

The relationship between the Fermi wavevectors defined above to the experimentally relevant bias voltage $V_b$ (measured relative to the semiconducting gap) and carrier density $n$ follow from the below relations (which are derived by considering the enclosed density of electrons/holes within the circular/elliptical Fermi surfaces):
\begin{align}
    n &= \frac{k_{F,v,x}k_{F,v,y}-k_{F,c,x}k_{F,c,y}}{4\pi} \nonumber\\
    V_b &= \frac{k_{F,v,x}^2}{2m_{v,xx}} + \frac{k_{F,c,x}^2}{2m_{c,xx}}
\end{align}
The Fermi momenta are related by $k_{F,v/c,y} = k_{F,v/c,x} \sqrt{m_{v/c,y}/m_{v/c,x}}$.
The momentum in all our plots is in units of $a^{-1}$.

\section{Density Wave Instability}
\label{app_dw}

At the non-interacting level, the system is described by an annualar Fermi surface, with an inner electron-like Fermi surface and an outer hole-like Fermi surface. Previous studies~\cite{pieri2007effects,varley2016} have established an instability of the annular Fermi surface to an interlayer charge density wave, which can be described by the order parameter $\< f\d_{v,\v k+\v Q} f_{c,\v k} \>$. We obtain the conditions required for such an instability for our model within mean-field theory. Performing a mean-field decoupling in the spiral X-DW channel of interest, we obtain the following mean-field Hamiltonian:
\begin{align}
    H_{\text{mf}} &= H^0_c + H^0_v + \Sigma \nonumber \\
    \Sigma &= -\frac{1}{A} \sum_{\v k, \v k'}  U(\v k' - \v k) f\d_{c,\v k} f_{v,\v k + \v Q} \< f\d_{v,\v k'+\v Q} f_{c,\v k'}\> + h.c.
\end{align}

Defining $\chi_{\v k} \equiv \frac{1}{A}\sum_{\v k'} U(\v k - \v k') \< f\d_{v,\v k'+\v Q} f_{c,\v k'}\>$, we can compactly write the mean-field Hamiltonian:
\begin{align}
    H_{\text{mf}} = \sum_{\v k} \begin{pmatrix}
        f\d_{c, \v k} &f\d_{v, \v k+\v Q}
    \end{pmatrix} \begin{pmatrix}
        \xi_{c, \v k} &-\chi_{\v k}\\
        -\chi^*_{\v k} &\xi_{v, \v k + \v Q}
    \end{pmatrix}\begin{pmatrix}
        f_{c, \v k} \\
        f_{v, \v k+\v Q}
    \end{pmatrix}
    \label{si:mf_ham}
\end{align}
Solving the mean-field Hamiltonian yields the following gap equation:
\begin{align}
    \chi_{\v k} = \frac{1}{A}\sum_{\v k'} V(\v k - \v k') \(n_F(\xi_{\v k'}-E_{\v k'}) - n_F(\xi_{\v k'}+E_{\v k'}) \)\frac{\chi_{\v k'}}{2E_{\v k'}}
    \label{eq_app:dw-gap}
\end{align}
where $\xi_{\v k} \equiv (\xi_{c,\v k} + \xi_{v,\v k+\v Q})/2$, $E_{\v k} \equiv \sqrt{(\xi_{c,\v k} - \xi_{v,\v k+\v Q})^2/4 + |\chi_{\v k}|^2}$ and $n_F$ is the Fermi-Dirac distribution function. Let us define the quasiparticle energies $\eps_{\pm, \v k} = \xi_{\v k} \pm E_{\v k}$ for future convenience. We work with a fixed total carrier density, which is set by a chemical potential term in the Hamiltonian.

This formulation to determine the exciton order parameter considers a select number of momentum shells/bands.
For the exciton density wave instabilities (both spiral and collinear X-DW), we hence also perform fully self-consistent Hartree-Fock theory, including up to 7 $\times$2 bands; the number of bands selected is such that the UV cutoff is reached.
We direct the reader to Ref. \cite{dong2023theory} for analogous Hartree-Fock calculations performed in rhombohedral stacked multilayer graphene.
We choose a UV cutoff of $\pm 0.1/a$ in the $\hat{y}$-direction, and the first Brillouin zone along the $\hat{x}$-direction bounded by the $\pm |{\v Q}|/2$.
The dielectric constant is taken to be $\epsilon=10$; typical choices of the hBN dielectric environment is $5 \sim 10$ to account for additional screening by the mobile charge carriers in a system with finite carrier density.
As described in the main text, our focus is on the inter-layer instabilities, rather than the intra-layer instabilities.
This amounts to suppressing the intra-layer Hartree and Fock contributions in the decoupling.

\begin{figure}[t]
\includegraphics[width = 0.4\textwidth]{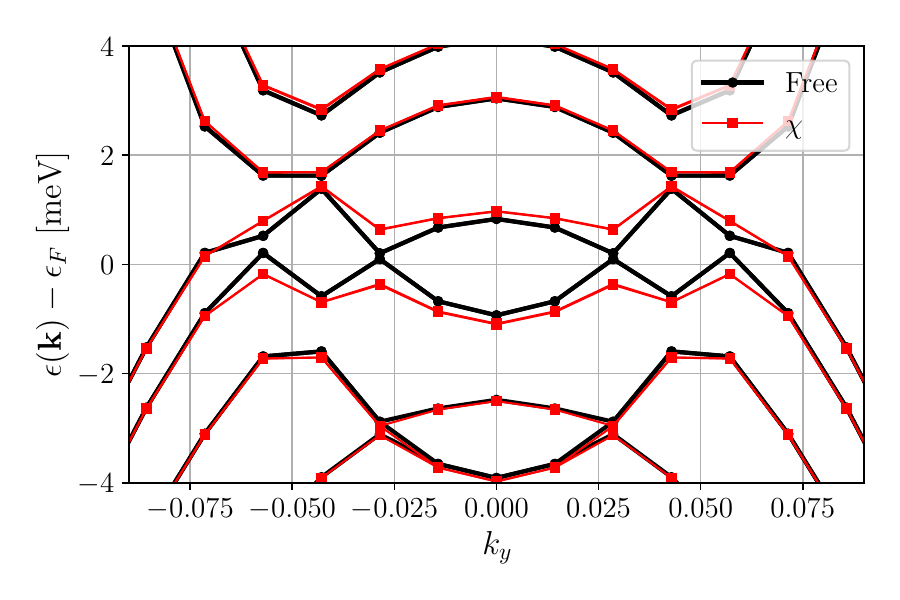}
\caption{Self-consistent electronic bandstructure at $k_x = - 0.00778$ for $n = - 0.26 \times 10^{12}$ cm$^{-2}$ and $V_b = 8.6$ meV.
The red/black colors indicate the bandstructure in the presence/absence of the collinear exciton condensate.
Momentum mesh size of 15$\times$15, and up-to 7$\times$2 (including both conduction and valence) folded bands was used.}
\label{fig_exciton_band_structure_app}
\end{figure}

We present in Fig. \ref{fig_exciton_band_structure_app} a representative electronic bandstructure depicting the free (black) and finite exciton density wave order parameter (red).
As seen, a electronic bands still cross the Fermi level, despite a finite exciton condensate, consistent with the three-band Hubbard approximation presented in Fig. 3 in Sec. III in the main text.

\section{Estimation of $V_{DW}$}
\label{si:rpa}

\begin{figure}[t]
\includegraphics[width = 0.4\columnwidth]{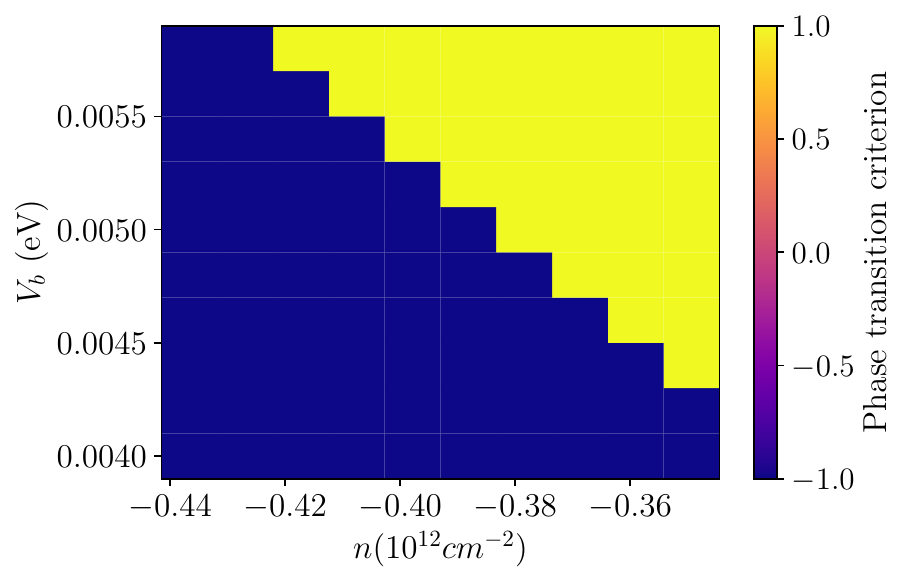}
\includegraphics[width = 0.4\columnwidth]{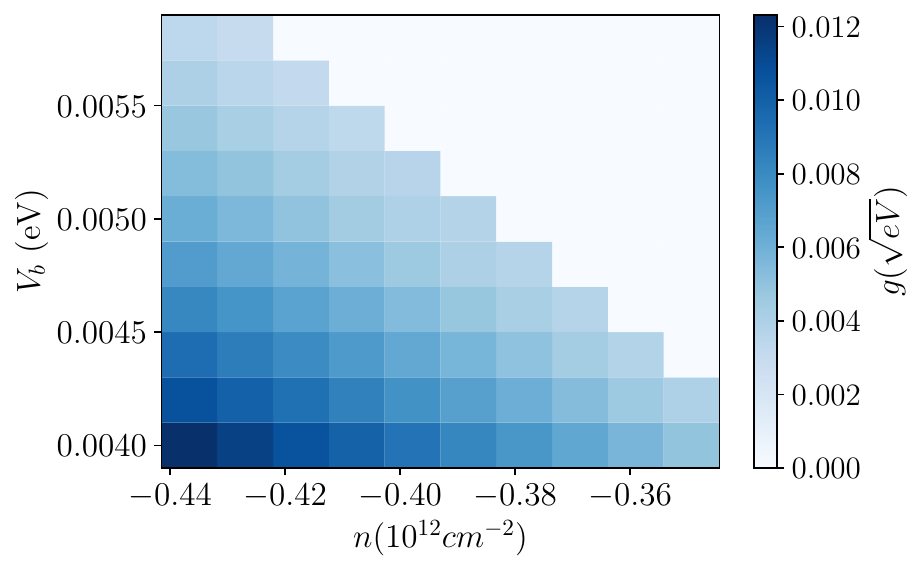}
\includegraphics[width = 0.4\columnwidth]{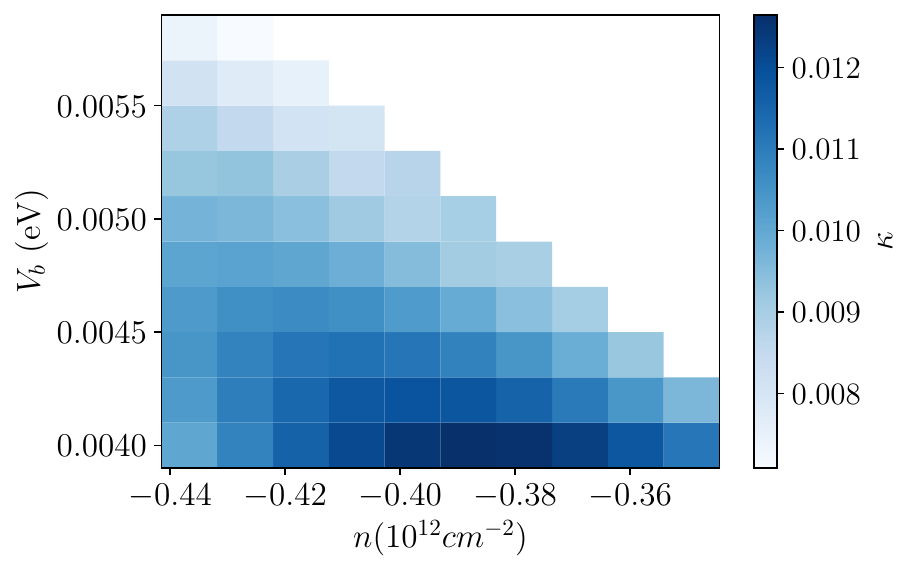}
\includegraphics[width = 0.4\columnwidth]{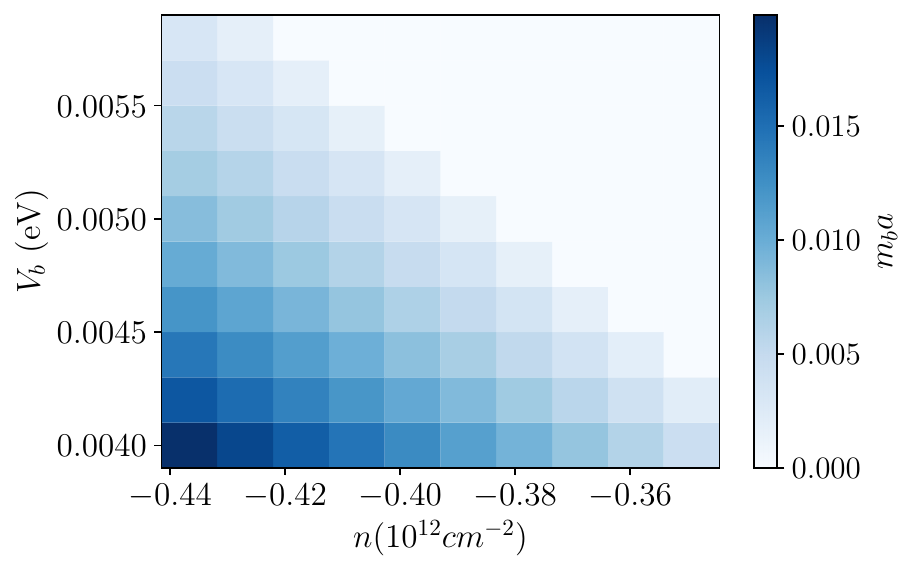}
\caption{Top left: The criterion for the X-DW phase transition within the Hubbard interaction approximation. Yellow and blue are respectively the symmetric ($\Pi_{cv}(\v Q)U_0 > -1$) and the X-DW ($\Pi_{cv}(\v Q)U_0 < -1$) phases. The other panels plot the interaction parameters $g, \kappa, m_b$ defined in $V_{DW}$ as a function of phase space parameters. Parameters used are: $\eps = 18$, $199 \times 199$ momentum mesh.}
\label{fig_int_parameters}
\end{figure}

In this section, we will provide a microscopic derivation of the interaction $V_{DW}$ mediated by the soft bosons, by assuming that the bare interactions are strongly gate-screened ($qd_g \ll 1$) to become Hubbard-like: 
\begin{align}
U_{sc}(\v q) \rightarrow U_0 = \frac{e^2 d_g}{2\eps_0\eps} 
\end{align}
We expect this approximation to be reasonable at small densities, therefore small Fermi momenta, such that the typical momenta are much smaller than $1/d_g$. Using the Hubbard interaction $U_0$, we can now compute the bubble and ladder sums (following Ref.~\cite{berk_schrieffer_1966, Scalapino1986}) to obtain the ``paramagnon" mediated interaction:
\begin{align}
    &\frac{1}{A}\sum_{\v k, \v k', \v q} V_{DW}(\v k, \v k', \v q) f\d_{v,\v k + \v q}f_{c,\v k} f\d_{c,\v k'}f_{v,\v k' + \v q} \\
    V_{DW}(\v k, \v k', \v q) &= -\frac{U_0^2 \Pi_{cv}(\v q)}{1+U_0 \Pi_{cv}(\v q)} + \frac{U_0}{1 - U_0 (\Pi_{cc}(\v k - \v k') + \Pi_{vv}(\v k - \v k'))}
\end{align}
where $\Pi_{cv}$ is the inter-band susceptibility and $\Pi_{cc/vv}$ are the corresponding intra-band susceptibilities. The first term is from the ladder sum and the second is due to the bubble sum (we have included the bare Hubbard interaction in the second term). The first term dominates near the DW-QCP and gets peaked near $\v q = \pm \v Q$, therefore, we ignore the second term in our analysis. We fit this interaction to the form specified in Eq. 8 of the main text to extract fit parameters $g,m_b$ and $\kappa$, as a function of $n$ and $V_b$. The fit parameters used in our calculations are plotted in Fig.~\ref{fig_int_parameters}.

\begin{figure}[t]
\includegraphics[width = 0.5\columnwidth]{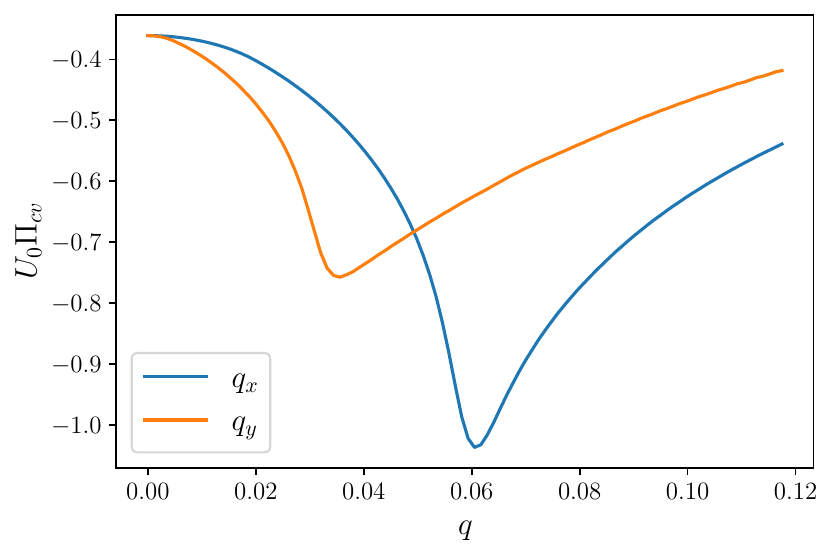}\\
\caption{$U_0 \Pi_{cv}$ as a function of momentum. As can be seen, $U_0 \Pi_{cv}$ along $\hat{x}$-direction is able to satisfy the density wave instability condition at smaller interaction strengths.}
\label{fig_susc_si}
\end{figure}

The density wave instability onsets when $U_0 \Pi_{cv}(\v q)=-1$ at momentum $\v q$. We present in Fig.~\ref{fig_susc_si} the behavior of $U_0 \Pi_{cv}(\v q)$ in the symmetric phase as a function of $\v q$ along the $x$ and $y$ directions.
As seen, the leading instability is for $\v q \sim \hat{x}$.
\textcolor{black}{The $\hat{x}$ direction is picked because of the larger effective mass of the valence band along that direction. If, instead, the system exhibited continuous rotational symmetry, all directions will be equally susceptible. Therefore, at the phase transition, there will be a circle in momentum space on which the exciton fluctuations become gapless. We anticipate the enhanced fluctuations from these gapless modes will turn the transition first order. We will hence focus on the discrete $C_{2z}$ symmetric case where the exciton modes become gapless only at two points. }

\section{Goldstone Mode Mediated Interactions}
\label{si:goldstone}
The two X-DW phases discussed in this work have gapless Goldstone modes due to breaking of continuous translation symmetry and the $U(1)$ interlayer charge conservation symmetry. The goal of this section is to study whether the interactions mediated by them can also lead to superconductivity. 
For analytical tractability of our analysis,
we assume that the gate-screened Coulomb interactions become over-screened in the limit of small gate distances $d_g \ll 1/q$, leading to a momentum-independent Hubbard interaction.

\subsection{Collinear X-DW}
The Hubbard-Stratonovich field associated with the X-DW couples locally to the fermions (in real space) as,
\begin{align}
    H_{\chi-f} = - \sum_{\bfr} \chi_{\bfr} f_{v, \bfr}^{\dag} f_{c, \bfr} + \text{h.c.}
\label{eq_spiral_dw_fermion_real_space_coupling}
\end{align}
Here $\chi_{\bfr} = \chi_0 \cos \v Q \cdot \v r + \delta \chi_{\v r} = \chi_0 \cos{\v Q \cdot \v r} + \delta \chi_-(\v r) e^{-i\v Q \cdot \v r}+ \delta \chi_+(\v r)e^{i \v Q \cdot \v r}$ where the collinear X-DW has a periodicity given by $\textbf{Q}$, $\chi_0$ is the uniform mean field exciton condensate, and $\delta \chi_\bfr$ is the fluctuation. 
For Hubbard interactions $U_0$, the corresponding momentum space mean-field order parameter 
is $\v k$-independent, and is self-consistently determined by solving the mean-field Hamiltonian.
We now focus on the fluctuating part of Eq. \ref{eq_spiral_dw_fermion_real_space_coupling} and rewrite it in momentum space as,
\begin{align}
    H_{\delta \chi-f} & = - \sum_{\bfq, \bfk} \delta \chi_-({-\bfq}) f_{c, \bfk }^{\dag} f_{v, \bfk - \textbf{Q}+\bfq} + \delta \chi_+(-{\bfq}) f_{c, \bfk }^{\dag} f_{v, \bfk + \textbf{Q}+\bfq} + \text{h.c.}
    \label{si:fermion_fluc}
\end{align}
Assuming that $\chi_0$ is real, we confine our focus to the examination of phase fluctuations about the mean-field order parameter, \textit{i.e.,} focus on the imaginary part of $\delta \chi_{\pm} = \delta \chi_{\pm,x} + i\delta \chi_{\pm,y}$.  

We now formally integrate out the (high-energy) fermions to obtain a fluctuation Hamiltonian:
\begin{align}
    H_{fluc.} = \frac{A}{2} \sum_{\v q} \delta \chi_{i,y}(-\v q) \(\frac{2}{U_0}\delta_{ij} - \Pi_{ij}(\v q, \Omega=0)\) \delta \chi_{j,y}(\v q).
    \label{eq_fluctation_action_d3}
\end{align}
Here $i,j$ run over the two fluctuations $\pm$, $A$ is the sample area, and $\delta_{ij}$ is the Kronecker delta function.
$\Pi_{ij}$ is calculated using the mean-field quasiparticle states. 
It is evident from the first term in Eq. \ref{eq_fluctation_action_d3} that we are employing the Hubbard approximation. 
The polarization bubble is given by,
\begin{align}
    \Pi_{ij}(\v q, \Omega) = -\int \frac{d^2 k d\omega}{(2\pi)^3} \text{Tr}\(\sigma_{i,y} G(\v k + \v q, \omega+\Omega) \sigma_{j,y} G(\v k,\omega)\),
\end{align}
where $G$ is the Green's function of the mean-field quasiparticles, and $\sigma_{\pm,y}$ is the Pauli matrix written in the basis $\begin{pmatrix}
        f\d_{c, \v k} &f\d_{v, \v k \pm \v Q}
    \end{pmatrix}^T$. 
Evaluating the above integral numerically, we obtain the following form, for small momentum.
\begin{align}
    H_{\text{fluc}} = \frac{1}{4}\sum_{\v q} \begin{pmatrix}
        \delta \chi_{-,y}(-\v q) &\delta \chi_{+,y}(-\v q)
    \end{pmatrix}\( (\rho^x_{s} + \rho^x_p)q_x^2 + (\rho^y_{s} + \rho^y_p)q_y^2 + \sigma_x((\rho^x_{s} - \rho^x_p)q_x^2 + (\rho^y_{s} - \rho^y_p)q_y^2) \)
    \begin{pmatrix}
        \delta \chi_{-,y}(\v q) \\ \delta \chi_{+,y}(\v q)
    \end{pmatrix}
    \label{app:eq_fluc}
\end{align}
The vanishing diagonal entries $\Pi_{ii}(\v q =0, \Omega=0)U_0 = 2$ and the off-diagonal entries $\Pi_{ij}(\v q =0, \Omega=0) = 0$ (for $i \neq j$) at zero momentum, ensuring a gapless spectrum.
Diagonalizing the fluctuation Hamiltonian, we obtain the gapless Goldstone modes $\delta \chi_{s/p}(\v q) = (\delta \chi_{-,y}(\v q) \pm \delta \chi_{+,y}(\v q))/\sqrt{2}$, and the fluctuation Hamiltonian in the main text (Eq. 4). 
In the presence of a lattice potential, $\delta \chi_s$ is a true Goldstone mode corresponding to the spontenous breaking of the interlayer charge conservation symmetry, whereas $\delta \chi_p$ stays gapless only if $\v Q$ is incommensurate with the lattice potential. 

Let us now calculate the coupling of these modes to the fermionic quasiparticles in the collinear X-DW phase. 
We define a Brillouin zone (BZ) along $\hat{x}$ direction due to the collinear X-DW ordering: $k_x \in [-Q,Q)$. For generic $(k_{F,c}$, $k_{F,v})$, the Fermi surface can be complicated due to bands originating in multiple zones that cross the Fermi level. Therefore, we focus on parameter points $(n, V_b)$ such that $k_{F,c,x} < Q$, which simplifies the Fermi surface to arise from only three bands: $f_{c,\v k}$, $f_{v,\v k + \v Q}$ and $f_{v,\v k - \v Q}$, where $\v k$ is defined in the BZ. 
For Hubbard interactions, the mean-field Hamiltonian is
\begin{align}
    H_{mf} = \sum_{\v k} \begin{pmatrix}
        f\d_{c, \v k} &f\d_{v, \v k+\v Q} &f\d_{v, \v k-\v Q}
    \end{pmatrix} \begin{pmatrix}
        \xi_{c, \v k} &-\chi_{+} &-\chi_-\\
        -\chi^*_{+} &\xi_{v, \v k + \v Q} &0 \\
        -\chi^*_- &0 &\xi_{v, \v k - \v Q}
    \end{pmatrix}\begin{pmatrix}
        f_{c, \v k} \\
        f_{v, \v k+\v Q}\\
        f_{v, \v k-\v Q}
    \end{pmatrix}
\end{align}
We numerically obtain real self-consistent solutions such that $\chi_+=\chi_-$, and find the wavefunction $\psi_{\v k}$ (defined in the 3-band basis) of the low-energy band $f_{\v k}$ that forms the Fermi surface with dispersion $\eps_{\v k}$. 
We can then obtain the fermion-Goldstone mode coupling between the low-energy fermions and the Goldstone modes:
\begin{blueblock}
\begin{align}
    \alpha_s(\v k', \v k) &= \psi_{\v k'}^*\begin{pmatrix}
        0 &-1 &-1\\
        1 &0 &0 \\
        1 &0 &0
    \end{pmatrix}\psi_{\v k} \nonumber \\
    \alpha_p(\v k', \v k) &= \psi_{\v k'}^*\begin{pmatrix}
        0 &-1 &1\\
        1 &0 &0 \\
        -1 &0 &0
    \end{pmatrix}\psi_{\v k} \nonumber \\
    H_{f-G} &= -i \int_{\v q, \v k}\alpha_G(\v k + \v q, \v k) \delta \chi_G(-\v q) f\d_{\v k + \v q} f_{\v k}.
\end{align}

    Writing $\psi_{\v k} = \begin{bmatrix}
        a_{\v k} &b_{\v k} &c_{\v k}
    \end{bmatrix}^T$, $C'_{2z}$ symmetry implies $\psi_{-\v k} = \begin{bmatrix}
        a_{\v k} &c_{\v k} &b_{\v k}
    \end{bmatrix}^T$. 
    Here, $a,b$ and $c$ are coefficients in our three-band basis, obtained by diagonalizing the mean-field Hamiltonian. They can be chosen to be real since $\chi_{\pm}$ are real.
    This implies that for the symmetric ($\delta \chi_+ = \delta \chi_-$) superfluid Goldstone mode $\alpha$ is an even function: $\alpha_s(\v k, \v k') = \alpha_s(-\v k, -\v k')$, and for the anti-symmetric ($\delta \chi_+ = -\delta \chi_-$) phonon mode, $\alpha$ is an an odd function: $\alpha_p(\v k, \v k') = -\alpha_p(-\v k, -\v k')$, as quoted in the main text.
\end{blueblock}
Finally, we integrate out the Goldstone modes to obtain an effective interaction term between the low-energy fermions:
\begin{align}
    \frac{1}{2A\rho_{G}}\sum_{\v k, \v k', \v q} f\d_{\v k + \v q} f_{\v k} f\d_{\v k' - \v q} f_{\v k'} \frac{\alpha_{G}(\v k + \v q, \v k) \alpha_{G}(\v k' - \v q, \v k')}{q_x^2+\kappa q_y^2}.
    \label{eq:goldstone_interaction}
\end{align}
We note that the strength of the interaction approaches a finite limit as $\v q \rightarrow 0$, analogous to the case of typical Goldstone modes such as acoustic phonons and gapless spin waves~\cite{Watanabe_2014}, since $\alpha(\v k + \v q,\v k)$ is linear in $\v q$ for small $\v q$.
Let us define the Goldstone mediated interactions strength:
\begin{align}
    V_G(\v k, \v k', \v q) = \frac{1}{\rho_G}\frac{\alpha_G(\v k + \v q, \v k) \alpha_G(\v k' - \v q, \v k')}{q_x^2 + \kappa_G q_y^2}.
\end{align}

\begin{figure}[t]
\includegraphics[width = 0.4\columnwidth]{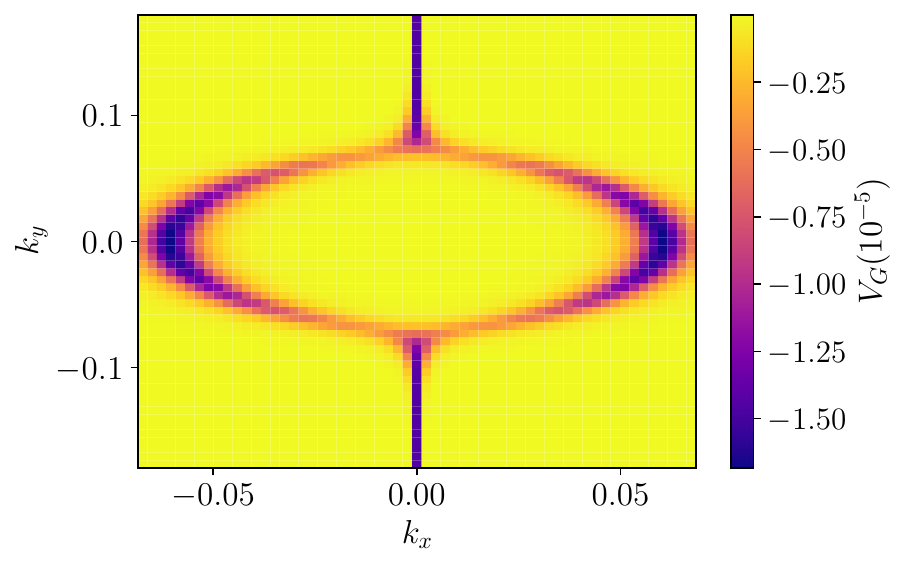}
\includegraphics[width = 0.4\columnwidth]{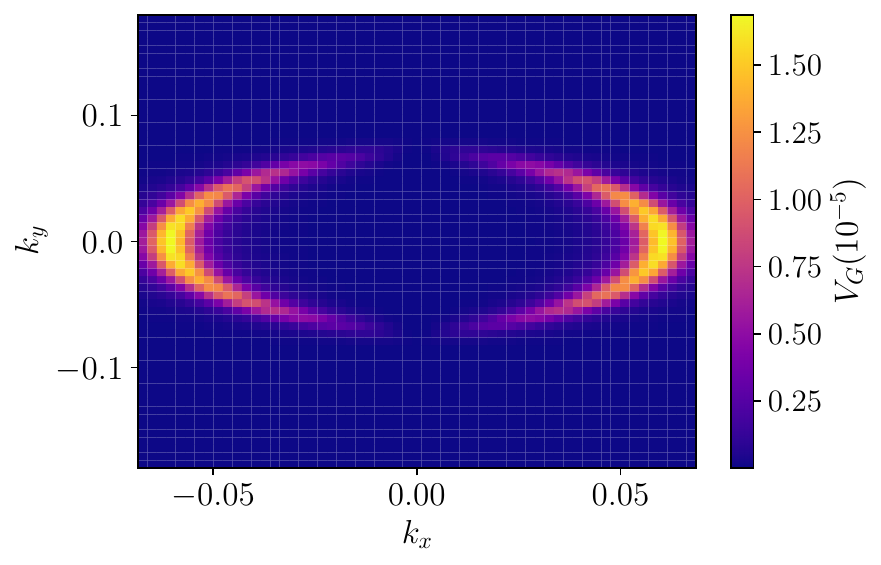}
\caption{Left/Right panel: Phonon/superfluid mode mediated interaction matrix elements for ${\v l} \rightarrow {\v k}$ in Eq. \ref{eq_d11}. The parameters used are $n=-0.7 \times 10^{12}$ cm$^{-2}$, $V_b=12$ meV and a $59 \times 59$ momentum mesh.}
\label{fig_collinear_goldstone_si}
\end{figure}

Having obtained the interactions mediated by the Goldstone modes, we now set up self-consistent BCS mean-field equations for superconductivity.
In the BCS channel, the interactions take the form
\begin{align}
    \frac{1}{2A}\sum_{\v k, \v l} f\d_{\v l} f\d_{- \v l} f_{-\v k} f_{\v k} V_G(\v k, -\v k, \v l - \v k).
    \label{eq_d11}
\end{align}
Let us define $V_{BCS}(\v k, \v l)=V_G(\v k,-\v k, \v l - \v k)/A$. We plot $V_{BCS}$ this interaction for $l_x \rightarrow k_x, l_y=k_y$ in Fig.~\ref{fig_collinear_goldstone_si}, to demonstrate the attractive/repulsive nature of the interactions mediated by $\delta \chi_{p/s}$. 
This motivates us to ask whether the Goldstone mode mediated interaction can lead to superconductivity. Decoupling the interaction term in the BCS channel, we get the following BdG Hamiltonian. 
\begin{align}
    H_{\text{BdG}} = \sum_{\v k} \begin{pmatrix}
        f\d_{\v k} &f_{-\v k}
    \end{pmatrix} \begin{pmatrix}
        \eps_{\v k} &\Delta_{BCS}(\v k)\\
        \Delta^*_{BCS}(\v k) &-\eps_{-\v k}
    \end{pmatrix}\begin{pmatrix}
        f_{\v k} \\
        f\d_{-\v k}
    \end{pmatrix}
\end{align}
where the superconducting gap:
\begin{align}
    \Delta_{BCS}(\v k) = \sum_{\v l} V_{BCS}(\v k, \v l) \< f_{-,-\v l}f_{-,\v l} \>
\end{align}
$\eps_{\v k} = \eps_{-\v k}$ due to the $C'_{2z}$ symmetry.
For a range of parameters, we find self-consistent superconducting solutions, as plotted in Fig. 4 in the main text.
The gap is maximized in the region where the interactions are strongly attractive.

\textcolor{black}{While we focus on $\v Q \neq 0$ states in this work, the above arguments let us draw conclusions about the (charge-imbalanced) $\v Q=0$ exciton condensed state as well. In this case, translational symmetry is not broken and therefore, there is only one Goldstone mode from the breaking of $U(1)_r$ symmetry: the superfluid mode $\delta \chi_s$. The fermion quasiparticles can be represented in terms of the two bands $f_{c,\v k}$ and $f_{v,\v k}$. The coupling between the Goldstone mode and the fermion quasiparticles, as shown for the superfluid mode above, is an even function of momentum in this case as well due to the $C'_{2z}$ symmetry. Following the discussion in the main text, this results in interactions that are repulsive and even-parity, ruling out superconductivity. }

\subsection{Spiral X-DW}
There is another possible exciton density wave phase where the system spontaneously breaks $C'_{2z}$ symmetry and picks either $\v Q$ or $-\v Q$ for the exciton ordering. In this ``spiral exciton density wave phase'', the phase of the exciton order parameter forms a spiral pattern. However, in our system, this is not the energetically preferred mean-field state.
In this phase, a combination of the generators remains as a symmetry,for example, the $U(1)_r$ phase transformation, $\chi = \chi_+ e^{i(\v Q \cdot \v r)} \rightarrow \chi_+ e^{i(\v Q \cdot \v r + \theta)}$, can be compensated by a lattice translation with $\alpha = -\theta$. This compensation entails the breaking of effectively one symmetry generator leading to one Goldstone mode.

For completness, we also examine the situation of Goldstone mediated superconductivity in the spiral X-DW.
The Hubbard-Stratonovich field associated with the sprial X-DW is $\chi_{\bfr} = e^{i \textbf{Q} \cdot \bfr} (\chi + \delta \chi_{\bfr})$ where the spiral X-DW has a periodicity given by $\textbf{Q}$, $\chi$ is the uniform mean field exciton condensate, and $\delta \chi_\bfr$ is the fluctuation. 
We note that we have extracted out the fast oscillatory behavior (arising from $\textbf{Q}$) explicitly in the order parameter.
The corresponding momentum space mean-field order parameter is self-consistently determined from the gap equation (which follows from Eq.~\ref{eq_app:dw-gap}),
\begin{align}
    \chi = \frac{U_0}{A}\sum_{\v k'} \(n_F(\eps_{-, \v k}) - n_F(\eps_{+, \v k}) \)\frac{\chi}{2E_{\v k'}}.
\end{align}
The fluctuations $\delta \chi$ couples to the fermions in the following way,
\begin{align}
    H_{\delta \chi-f} & = - \sum_{\bfq, \bfk} \delta \chi_{-\bfq} f_{c, \bfk }^{\dag} f_{v, \bfk + \textbf{Q}+\bfq} + \text{h.c.} \nonumber \\
    & = - \sum_{\bfq} \Big[ \delta \chi_x(-\v q) S_x(\v q) + \delta \chi_y(-\v q) S_y(\v q) \Big] \nonumber \\
    & = - \sum_{\bfq, \bfk} \vec{f}_{\bfq, \bfk}^{ \ \dag} \Big[ \vec{\delta \chi}({-\bfq}) \cdot \vec{\sigma} \Big] \vec{f}_{\bfq, \bfk}
    \label{si:fermion_fluc},
\end{align}
where we have used the fact that $\delta \chi_r = \delta \chi_x (\bfr) + i \delta \chi_y (\bfr)$; this implies that $[ \delta \chi_{x/y}(\bfq) ]^* = \delta \chi_{x/y}(-\bfq)$, and $S_x(\v q) \equiv \sum_{\v k} \(f\d_{v, \v k+ \v q+ \v Q}f_{c, \v k} + f\d_{c, \v k }f_{v, \v k + \v q + \v Q}\)$, $S_y(\v q) \equiv -i \sum_{\v k} \(-f\d_{v, \v k+ \v q+ \v Q}f_{c, \v k} + f\d_{c, \v k }f_{v, \v k + \v q + \v Q}\)$, and $\vec{f}_{\bfq, \bfk} = (f_{c, \bfk} \ f_{v, \v k + \v q + \v Q})^{\text{T}}$ where we express the fermionic operators in the basis of the mean-field Hamiltonian Eq. \ref{si:mf_ham}.

Integrating out high-energy fermions, we obtain
\begin{align}
    H_{fluc.} = \frac{A}{2} \sum_{\v q} \delta \chi_i(-\v q) \(\frac{2}{U_0}\delta_{ij} - \Pi_{ij}(\v q, \Omega=0)\) \delta \chi_j(\v q).
\end{align}
Here $i,j$ run over $x$ and $y$ and $\delta_{ij}$ is the Kronecker delta function.
$\Pi_{ij}$ is calculated using the mean-field quasiparticle states. 
\begin{align}
    \Pi_{ij}(\v q, \Omega) = -\int \frac{d^2 k d\omega}{(2\pi)^3} \text{Tr}\(\sigma_i G(\v k + \v q, \omega+\Omega) \sigma_j G(\v k,\omega)\)
\end{align}
where $G$ is the Green's function of the mean-field quasiparticles and $\sigma$ are the Pauli matrices written in the basis $\begin{pmatrix}
        f\d_{c, \v k} &f\d_{v, \v k+\v Q}
    \end{pmatrix}^T$. 

Carrying out the integral over $\omega$ for the static case $\Omega=0$, we obtain
\begin{align}
    \Pi_{xx/yy}(\v q) = &-\int \frac{d^2k}{(2\pi)^2} \frac{\mp 2\chi^2 - (\eps_{+,\v k + \v q} - \xi_{v, \v k + \v Q + \v q})(\eps_{+,\v k + \v q} - \xi_{c, \v k}) - (\eps_{+, \v k + \v q} - \xi_{c, \v k + \v q})(\eps_{+, \v k + \v q} - \xi_{v, \v k + \v Q})}{(\eps_{+, \v k + \v q} - \eps_{-, \v k + \v q})(\eps_{+, \v k + \v q} - \eps_{+, \v k})(\eps_{+, \v k + \v q} - \eps_{-, \v k})}\theta(\eps_{+,\v k + \v q}) \nonumber \\
    &-\int \frac{d^2k}{(2\pi)^2} \frac{\mp 2\chi^2 - (\eps_{-,\v k + \v q} - \xi_{v, \v k + \v Q + \v q})(\eps_{-,\v k + \v q} - \xi_{c, \v k}) - (\eps_{-, \v k + \v q} - \xi_{c, \v k + \v q})(\eps_{-, \v k + \v q} - \xi_{v, \v k + \v Q})}{(\eps_{-, \v k + \v q} - \eps_{+, \v k + \v q})(\eps_{-, \v k + \v q} - \eps_{+, \v k})(\eps_{-, \v k + \v q} - \eps_{-, \v k})}\theta(\eps_{-,\v k + \v q}) \nonumber \\
    &-\int \frac{d^2k}{(2\pi)^2} \frac{\mp 2\chi^2 - (\eps_{+,\v k} - \xi_{v, \v k + \v Q + \v q})(\eps_{+,\v k} - \xi_{c, \v k}) - (\eps_{+, \v k} - \xi_{c, \v k + \v q})(\eps_{+, \v k} - \xi_{v, \v k + \v Q})}{(\eps_{+, \v k} - \eps_{+, \v k + \v q})(\eps_{+, \v k} - \eps_{-, \v k + \v q})(\eps_{+, \v k} - \eps_{-, \v k})}\theta(\eps_{+,\v k}) \nonumber \\
    &-\int \frac{d^2k}{(2\pi)^2} \frac{\mp 2\chi^2 - (\eps_{-,\v k} - \xi_{v, \v k + \v Q + \v q})(\eps_{-,\v k} - \xi_{c, \v k}) - (\eps_{-, \v k} - \xi_{c, \v k + \v q})(\eps_{-, \v k} - \xi_{v, \v k + \v Q})}{(\eps_{-, \v k} - \eps_{+, \v k + \v q})(\eps_{-, \v k} - \eps_{-, \v k + \v q})(\eps_{-, \v k} - \eps_{+, \v k})}\theta(\eps_{-,\v k})\nonumber \\
    &=\Pi_{xx/yy}(-\v q)
\end{align}
and 
\begin{align}
    -i\Pi_{xy}(\v q) = &-\int \frac{d^2k}{(2\pi)^2} \frac{ (\eps_{+,\v k + \v q} - \xi_{v, \v k + \v Q + \v q})(\eps_{+,\v k + \v q} - \xi_{c, \v k}) - (\eps_{+, \v k + \v q} - \xi_{c, \v k + \v q})(\eps_{+, \v k + \v q} - \xi_{v, \v k + \v Q})}{(\eps_{+, \v k + \v q} - \eps_{-, \v k + \v q})(\eps_{+, \v k + \v q} - \eps_{+, \v k})(\eps_{+, \v k + \v q} - \eps_{-, \v k})}\theta(\eps_{+,\v k + \v q}) \nonumber \\
    &-\int \frac{d^2k}{(2\pi)^2} \frac{(\eps_{-,\v k + \v q} - \xi_{v, \v k + \v Q + \v q})(\eps_{-,\v k + \v q} - \xi_{c, \v k}) - (\eps_{-, \v k + \v q} - \xi_{c, \v k + \v q})(\eps_{-, \v k + \v q} - \xi_{v, \v k + \v Q})}{(\eps_{-, \v k + \v q} - \eps_{+, \v k + \v q})(\eps_{-, \v k + \v q} - \eps_{+, \v k})(\eps_{-, \v k + \v q} - \eps_{-, \v k})}\theta(\eps_{-,\v k + \v q}) \nonumber \\
    &-\int \frac{d^2k}{(2\pi)^2} \frac{(\eps_{+,\v k} - \xi_{v, \v k + \v Q + \v q})(\eps_{+,\v k} - \xi_{c, \v k}) - (\eps_{+, \v k} - \xi_{c, \v k + \v q})(\eps_{+, \v k} - \xi_{v, \v k + \v Q})}{(\eps_{+, \v k} - \eps_{+, \v k + \v q})(\eps_{+, \v k} - \eps_{-, \v k + \v q})(\eps_{+, \v k} - \eps_{-, \v k})}\theta(\eps_{+,\v k}) \nonumber \\
    &-\int \frac{d^2k}{(2\pi)^2} \frac{(\eps_{-,\v k} - \xi_{v, \v k + \v Q + \v q})(\eps_{-,\v k} - \xi_{c, \v k}) - (\eps_{-, \v k} - \xi_{c, \v k + \v q})(\eps_{-, \v k} - \xi_{v, \v k + \v Q})}{(\eps_{-, \v k} - \eps_{+, \v k + \v q})(\eps_{-, \v k} - \eps_{-, \v k + \v q})(\eps_{-, \v k} - \eps_{+, \v k})}\theta(\eps_{-,\v k}) \nonumber \\
    &= -i\Pi_{yx}(-\v q) \nonumber \\
    &= i\Pi_{xy}(-\v q)
\end{align}

We numerically evaluate the above integrals and find the following functional form for small momentum.
\begin{align}
    H_{fluc.} = \sum_{\v q} \begin{pmatrix}
        \delta \chi_x(-\v q) &\delta \chi_y(-\v q)
    \end{pmatrix}\begin{pmatrix}
        \rho^x_{xx}q_x^2 + \rho^y_{xx}q_y^2 + m_{xx} &is_{xy}q_x \\
        -is_{xy}q_x &\rho^x_{yy}q_x^2 + \rho^y_{yy}q_y^2
    \end{pmatrix}
    \begin{pmatrix}
        \delta \chi_x(\v q) \\ \delta \chi_y(\v q)
    \end{pmatrix}
    \label{app:eq_fluc}
\end{align}
$\rho_{xx/yy}$, $s_{xy}$ and $m_{xx}$ are fit parameters obtained numerically. 
The form of Eq.~\ref{app:eq_fluc} follows from the even (odd)-in-momentum property of the diagonal (off-diagonal) elements of $\Pi_{\alpha \beta}(\v q)$. 

In particular, we find that $U_0\Pi_{yy}(\v q) \rightarrow 2$ as $\v q \rightarrow 0$, which ensures the gaplessness of the Goldstone mode.
At $\v q=0$, the $x$ (``amplitude'') and $y$ (``phase'') fluctuations decouple, and the Goldstone mode is a purely phase fluctuation; we note the contrast to the $x$ (amplitude) mode which is gapped due to the mass term $m_{xx}$.
However, for $\v q \neq 0$, the collective modes are linear combinations of the amplitude and the phase fluctuations, which we obtain by diagonalizing the fluctuation action. 
The subsequently named Goldstone mode corresponds to the eigenvalue of the action that is gapless in nature.
The Goldstone mode can be expressed as a linear combination of the afore-defined fluctuation components $\delta_{x/y}$ as,
\begin{align}
    \delta \chi^*_G(\v q) &= i \sin \(\frac{X(\v q)}{2}\) \delta \chi^*_x(\v q) + \cos \(\frac{X(\v q)}{2}\) \delta \chi^*_y(\v q) \nonumber \\
    \sin X(\v q) &= \frac{-s_{xy}q_x}{\sqrt{\((\rho^x_{xx}-\rho^x_{yy})q_x^2 + (\rho^y_{xx}-\rho^y_{yy})q_y^2 + m_{xx}\)^2/4 + s_{xy}^2 q_x^2}}
\end{align}

\begin{figure}[t]
\includegraphics[width = 0.4\columnwidth]{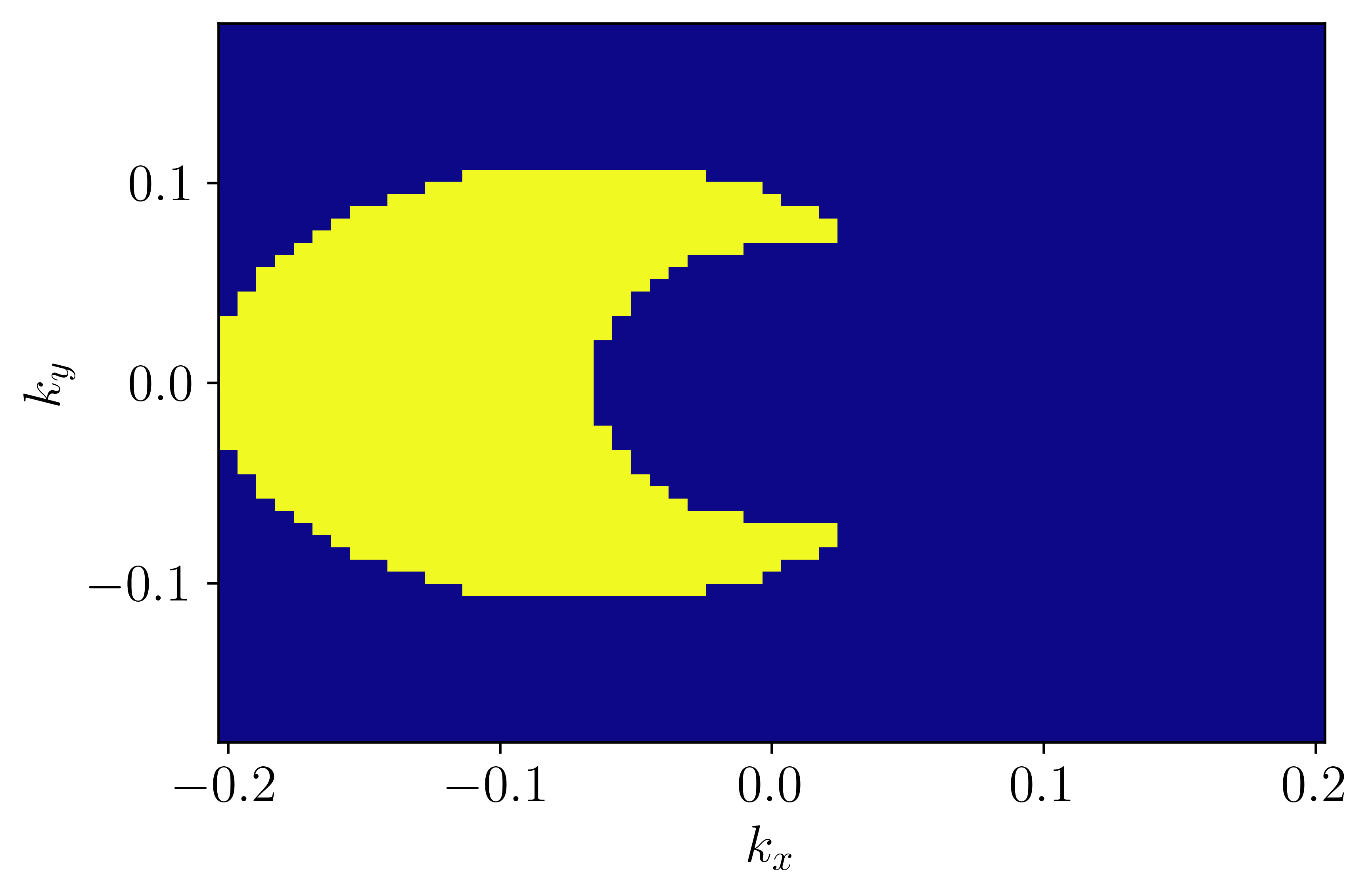}
\includegraphics[width = 0.4\columnwidth]{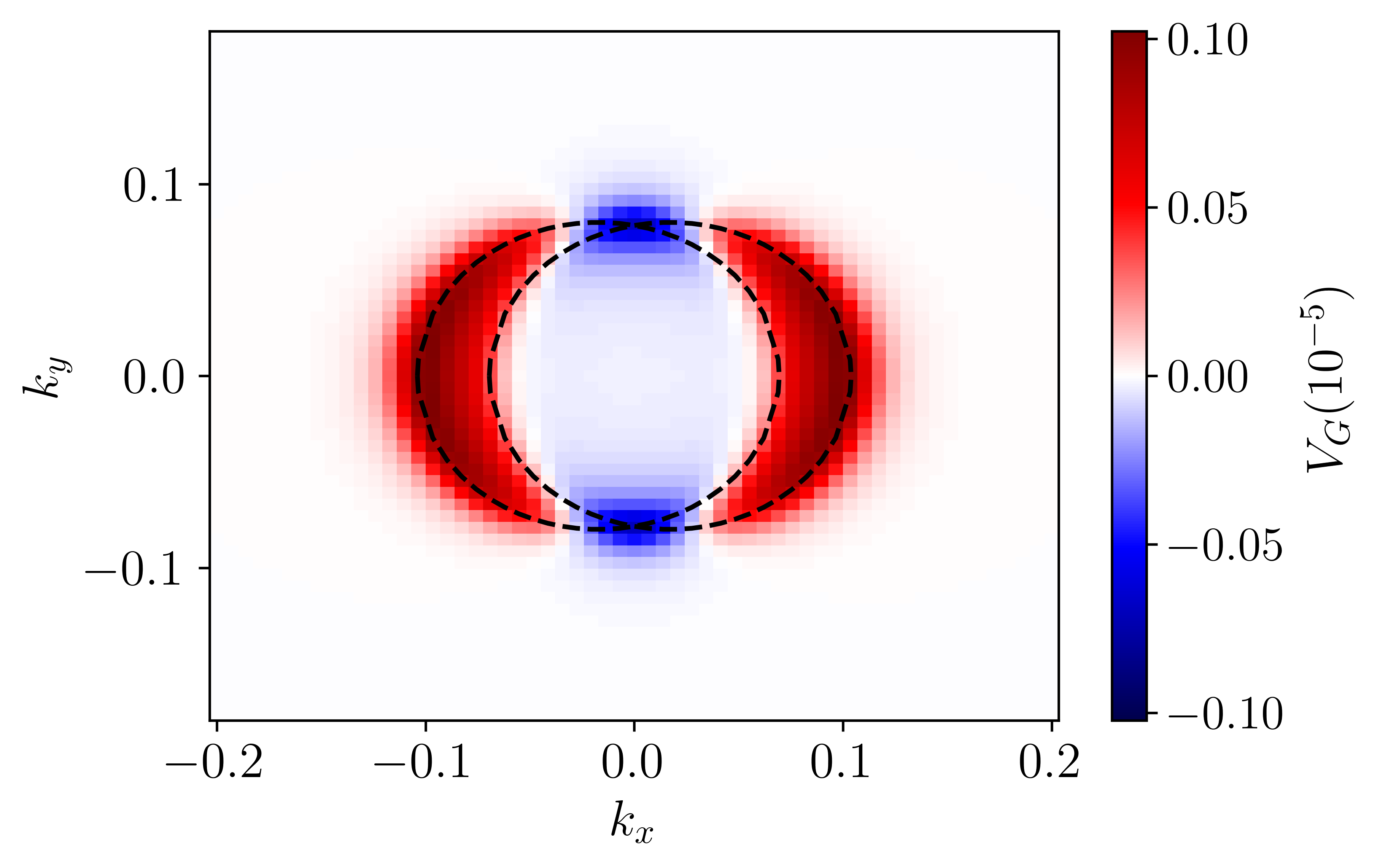}
\caption{Spiral X-DW. Left panel: Fermi surface of the spiral X-DW with yellow marking occupied states. Right panel: Goldstone mode mediated interaction matrix elements. Dashed lines plot the locus of points satisfying $\xi_{c,\v k}=\xi_{v,\v k \pm \v Q}$ (curve to the left/right). The parameters used are $n=-0.7 \times 10^{12}$ cm$^{-2}$, $V_b=14.4$ meV and a $59 \times 59$ momentum mesh.}
\label{fig_goldstone_si}
\end{figure}

Our next step is to calculate the interactions between the collective modes and the low-energy fermionic quasiparticles, as done for the collear case. We shall focus on the Goldstone fluctuations in the following, since the gapped collective mode would mediate relatively weaker interactions between the fermions away from the quantum critical line. 
Since we are considering hole doping, the Fermi surface is formed by the lower band $f_-$, which is obtained by diagonalizing the mean-field Hamiltonian~\ref{si:mf_ham} within the Hubbard approximation to yield eigenmodes,
\begin{align}
    f_{-, \v k} = -\sin \( \frac{\theta(\v k)}{2}\) f_{c,\v k} + \cos \( \frac{\theta(\v k)}{2}\) f_{v,\v k + \v Q},
\end{align}
where $\sin \theta(\v k) = -\frac{\chi}{E_{\v k}}$. 
We can thus rewrite Eq.~\ref{si:fermion_fluc} in terms of the Goldstone mode and lower-band fermion operators to obtain the 
spin-independent) fermion-Goldstone mode interaction $\alpha$, where the form factor is:
\begin{align}
    \alpha(\v k + \v q, \v k) = \cos \(\frac{X(\v q)}{2}\) \sin \(\frac{\theta(\v k + \v q) - \theta(\v k)}{2}\) + \sin \(\frac{X(\v q)}{2}\) \sin \(\frac{\theta(\v k + \v q) + \theta(\v k)}{2}\) .
\end{align}
Finally, integrating out the Goldstone mode, we obtain an effective four-fermion interaction term identical to the one derived for the collinear case in Eq.\ref{eq:goldstone_interaction}.
Fig.\ref{fig_goldstone_si} (right panel) plots the interaction $V_G$ as $l_x \rightarrow k_x$, $l_y=k_y$.
The interaction is attractive in a region of momentum space. However, since the spiral X-DW state breaks $C'_{2z}$, states at momenta $\v k$ and $-\v k$ are not degenerate, and it would require strong attractive interactions for them to overcome the energy difference and pair. While such a superconducting instability can occur in principle, our mean-field calculations for superconductivity do not yield superconducting solutions in the parameter space explored in this work. 

\section{Superconductivity Mediated by Critical Fluctuations}
\label{app_sc_pdw}
\begin{figure}[t]
\includegraphics[width = 0.4\columnwidth]{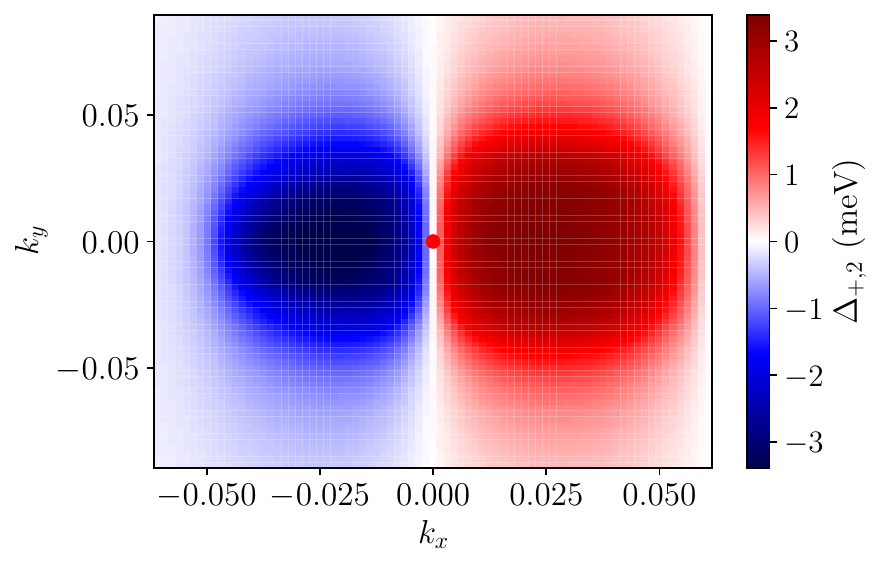}
\includegraphics[width = 0.4\columnwidth]{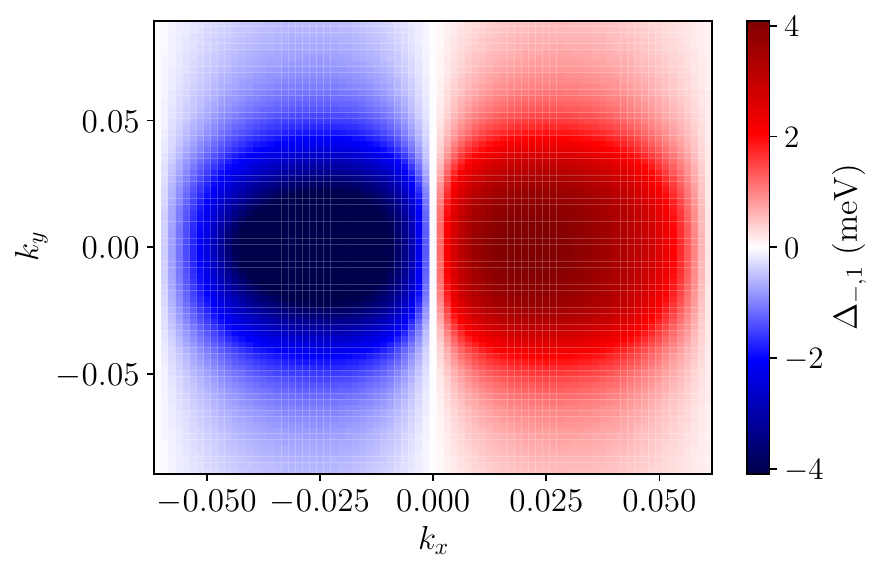}
\caption{Quantum critical fluctuation-mediated superconductivity. Pair-density wave (PDW) gap for $n=-0.36\times 10^{12}$ cm$^{-2}$ and $V_b = 4.2$ meV obtained using a $79 \times 79$ momentum mesh.}
\label{fig_pdw_si}
\end{figure}

For Cooper pairs having a center-of-mass momentum $\v Q$, we decompose the interactions in Eq. 5 of the main text in the PDW channel:
\begin{align}
    \frac{1}{A} \sum_{\v k, \v k'} V_{DW}(-\v k - \v k' + \v Q)f\d_{c,\v k} f\d_{v,-\v k + \v Q} f_{v, -\v k'+\v Q} f_{c,  \v k'} + V_{DW}(-\v k - \v k' - \v Q)f\d_{c,\v k} f\d_{v,-\v k - \v Q} f_{v, -\v k'-\v Q} f_{c,  \v k'} 
\end{align}
The finite-momentum Cooper pair condensation introduces a periodicity along the $x$-direction defined by the Brillouin zone $k_x \in [-Q_x, Q_x)$, with the $k_y$ momentum still defined in the continuum.
Let us define PDW gap functions:
\begin{align}
    \Delta_{\pm,n}(\v k) &= \frac{1}{A} \sum_{\v k',m} V_{DW}(-\v k - \v k'\pm (1-2n)\v Q- 2m \v Q) \< f_{v, -\v k' \pm \v Q - 2m \v Q} f_{c, \v k'+2m \v Q} \> .
\end{align}
where $n$ is a non-negative integer, and $m$ runs over all integers. In our calculations we keep five lowest energy states: $f_{c,\v k}, f_{c,\v k \pm \v Q}$, $f_{v,-\v k \pm \v Q}$. The BdG Hamiltonian is then
\begin{align}
    H_{\text{BdG}} = \sum_{\v k} \begin{pmatrix}
        f\d_{c, \v k+2\v Q} &f_{v, -\v k + \v Q} &f\d_{c, \v k}  &f_{v, -\v k - \v Q} &f\d_{c, \v k-2\v Q} 
    \end{pmatrix} \begin{pmatrix}
        \xi_{c, \v k - 2\v Q} &0 &0 &\Delta_{+,2}(\v k) &0\\
        0 &-\xi_{v, -\v k + \v Q} &\Delta_{+,1}(\v k) &0 &\Delta_{-,2}(\v k)\\
        0 &\Delta^*_{+,1}(\v k) &\xi_{c,\v k} &\Delta_{-,1}(\v k) &0 \\
        \Delta^*_{+,2}(\v k) &0 &\Delta^*_{-,1}(\v k) &-\xi_{v,-\v k - \v Q} &0 \\
        0 &\Delta^*_{-,2}(\v k) &0 &0 &\xi_{c, \v k + 2\v Q}
    \end{pmatrix}\begin{pmatrix}
        f_{c, \v k+2\v Q} \\
        f\d_{v, -\v k + \v Q} \\
        f_{c, \v k} \\
        f\d_{v, -\v k - \v Q} \\
        f_{c, \v k - 2\v Q}
    \end{pmatrix}.
\end{align}
We now numerically diagonalize $H_{\text{BdG}}$ and solve for the gap functions self-consistently. 
For our calculations, we truncate to keep the five lowest energy bands and obtain $\Delta_+ = \Delta_-$.
Fig.~\ref{fig_pdw_si} plots the obtained $\Delta_{\pm,1}$ as a function of $\v k$. 

We note that we also considered the possibility of zero-momentum BCS pairing channel and follow an analogous gap equation approach. 
In this case, the interaction term is decomposed in the BCS channel
\begin{align}
    \frac{1}{A}\sum_{\v k, \v k'} V_{DW}(-\v k - \v k') f\d_{c,\v k} f\d_{v,-\v k} f_{v, -\v k'} f_{c, \v k'}.
\end{align}
In the parameter space considered in this work, we do not find non-trivial self-consistent solutions for the order parameter $\< f_{v,-\v k} f_{c,\v k} \>$.

\section{NFL Properties}
\label{si:nfl}
Here we describe the properties of the NFL within a one-loop calculation. A standard approach is to study low-energy fermionic excitations near certain points on the Fermi surface, referred to as hot-spots. The fermions here are expected to be affected most severely by the coupling to the quantum critical fluctuations. 

Let us first consider points connected by momentum $\v Q$. Electron dispersion at these patches is (linearized at low-energies, with momenta defined relative to the hot-spots): $\eps_c(\v k) = v_c k_x + \frac{k_y^2}{2m_c}$ and $\eps_v(\v k) = -v_v k_x - \frac{k_y^2}{2m_v}$, where the Fermi velocities are defined to be positive: $v_{c/v}>0$. The inter-band polarizability for these points is given by:
\begin{align}
    \Pi_{cv}(\v q, \Omega) = \int \frac{d^2k d\omega}{(2\pi)^3} \frac{1}{i(\omega+\Omega)-v_c(k_x+q_x)-\frac{(k_y+q_y)^2}{2m_c}} \frac{1}{i\omega+v_v k_x + \frac{k_y^2}{2m_v}}.
\end{align}
Low-energy electron scatterings correspond to $q_x=0$ and $q_y \neq 0$, therefore, we set $q_x=0$, in order to extract the leading singular contribution. Let us also assume $m_c=m_v=m$ 
, which will not affect our conclusions qualitatively\footnote{We generically consider $v_c \neq v_v$. For the case $v_c = v_v$, we note that $\Pi_{cv}$ exhibits a logarithmic IR divergence in the static limit within our model: $\Omega(\v q, \Omega=0) \sim \log |q_y|$.}.

Evaluating first the $k_x$ integral, we obtain
\begin{align}
    \Pi_{cv}(\v q,\Omega) = -i\int \frac{dk_y d\omega}{(2\pi)^2} \frac{\theta(\omega+\Omega) - \theta(-\omega)}{i((v_v+v_c)\omega + v_v\Omega) + v_c \frac{k_y^2}{2m} - v_v\frac{(k_y+q_y)^2}{2m}}.
\end{align}
Next, we perform the $\omega$ integral by introducing a UV cutoff $\Lambda$. 
\begin{align}
    \Pi_{cv}(\v q, \Omega) = \frac{1}{2\pi(v_c + v_v)} \int \frac{dk_y}{2\pi} \log \( -i\Omega v_c + v_c\frac{k_y^2}{2m} - v_v\frac{(k_y + q_y)^2}{2m} \) + \log \( i\Omega v_v + v_c\frac{k_y^2}{2m} - v_v\frac{(k_y + q_y)^2}{2m} \)
\end{align}

Retaining the singular terms (as $\v q \rightarrow 0$) in the $k_y$ integral results in,
\begin{align}
    \Pi_{cv}(\v q, \Omega) \approx 
    \frac{m}{2 \pi (v_c + v_v)} \left( \sqrt{\frac{v_c}{v_v}}+ \sqrt{\frac{v_v}{v_c}} \right) \frac{ | \Omega |}{|q_y|},
\end{align}
where we have dropped non-singular cutoff-$\Lambda$ dependent terms.
The one-loop boson self-energy is then $g^2 \Pi_{cv}(\v q, \Omega)$. Adding this to the boson action, we obtain the quantum critical ($m_b=0$) action
\begin{align}
    S_b = \sum_{s}\int \frac{d^2q d\Omega}{(2\pi)^3} \(\Omega^2 + (\v q + s \v Q)^2  + \frac{mg^2}{2 \pi (v_c + v_v)} \left( \sqrt{\frac{v_c}{v_v}}+ \sqrt{\frac{v_v}{v_c}} \right) \frac{ | \Omega |}{|q_y|} \) |b_s(\v q, \Omega)|^2 .
\end{align}
This yields a dynamical critical exponent of $z=3$, which leads to  specific heat scaling at low temperatures as $C \sim T^{2/3}$~\cite{sslee_review_nfl_2018}.
Using the modified boson propagator, we calculate the one-loop fermion self-energy. The valence electron self-energy is,
\begin{align}
    \Sigma_v(\v k, \omega) = g^2 \int \frac{d^2q d\Omega}{(2\pi)^3} \frac{1}{q_y^2 + \frac{mg^2}{2 \pi (v_c + v_v)} \left( \sqrt{\frac{v_c}{v_v}}+ \sqrt{\frac{v_v}{v_c}} \right) \frac{ | \Omega |}{|q_y|}} G_c(\v k + \v q, \omega + \Omega),
\end{align}
where $G_c$ is the Green's function of the conduction fermions. For fermions close to the Fermi surface, we obtain
\begin{align}
    \Sigma_v(\omega) \sim i|\omega|^{2/3} \text{sgn}(\omega)
\end{align}
indicating a breakdown of Fermi liquid quasiparticles near the hotspots. We comment that the self-energies obtained here are qualitatively the same as the problem of a Fermi surface coupled to a $\v Q=0$ critical boson order parameter \cite{Sachdev_2011}. In this case, the Landau quasiparticles get destabilized everywhere on the Fermi surface, but in contrast, in our case, only the fermions near the hot-spots are affected. Note also that the answers obtained here are distinct from the commonly analyzed case of spin-density wave criticality in cuprates, where the boson self-energy is linear in $|\Omega|$ with a constant $q$-independent denominator. This further leads to deviations in the fermionic self-energy as well. The main qualitative difference is that the Fermi velocities at the hot-spots connected by the ordering wave-vector are non-collinear in the SDW case, whereas here, they are collinear.

Motivated by previous work~\cite{senthil2014massenhancementnearoptimal} on diverging quantum oscillation mass near a quantum phase transition, we analyze this possibility in the present context. 
The quantum oscillation mass $m_{\text{QO}}$ is expressed as a line integral over the Fermi surface: $m_{\text{QO}} = \frac{1}{2\pi} \oint \frac{dK}{v_{F}(\theta)}$, where $\theta$ labels Fermi surface patches, and $v_F(\theta)$ is the Fermi velocity. Let us approach the phase transition to the X-DW phase from the electron-hole plasma state where there is no exciton ordering. Here, we expect a Fermi liquid form for the fermion Green's function:
\begin{align}
    G(\v k, \omega) = \frac{Z}{i\omega - v_{F}(\theta)k_{\perp}}.
    \label{eq_app:green}
\end{align}
Near the hot-spots and close to criticality, we anticipate the following scaling form for the Green's function: $G \sim g(\frac{\omega}{k_{\parallel}^z}, \frac{k_{\perp}}{k_{\parallel}^2}, \frac{\omega}{\delta})$, where $\delta$ is an energy scale that characterizes the distance to the quantum critical point, $k_{\perp}$ ($k_{\parallel}$) quantifies the momentum deviation from the hotspots along the normal (tangential) direction to the Fermi surface. Comparing with the Fermi liquid form in Eq.~\ref{eq_app:green}, we obtain $v_F \sim |k_{\parallel}|^{z-2}$. Since $z=3$ in our case (on approach to the quantum critical point) we expect that $v_F \sim |k_{\parallel}|$ and thus vanishes at the hot spot.
Therefore, $m_{\text{QO}} \sim \oint dk_{\parallel}/k_{\parallel}$ diverges logarithmically.

We can however ask how $m_{\text{QO}}$ diverges as we approach the quantum critical point. 
Consider tuning the bias voltage $V_b$ towards a critical value $V_b^*$. Then, we have $\delta \sim |V_b - V_b^*|^{\nu z}$, where $\nu$ is a universal critical exponent. 
The integral over $k_{\parallel}$ is now cut-off by a finite momentum scale $\delta^{1/z}$ which leads to,
\begin{align}
    m_{\text{QO}} \sim \log \delta \sim \log |V_b - V_b^*|,
\end{align}
This logarithmic scaling could be detected in Shubnikov-de Haas measurements.

\section{Linear Response}
\label{si:response}
In the spiral X-DW phase, we can write $\chi \sim e^{i\phi}$ from which the low-energy Lagrangian can be written as: $\frac{\rho_G}{2}\(\partial_{\mu} \phi - \v Q\)^2$, so as to favor spiral ordering $\phi = \v Q \cdot \v r$.
Let us now introduce electromagnetic gauge fields in each of the layers $A_{c/v}$, assuming that the two layers have been separately contacted. 
We then have,
\begin{align}
    \L_s[\phi,A_c,A_v] = \frac{\rho_G}{2} \( \partial_{\mu} \phi - \v Q - (A_{c,\mu}-A_{v,\mu})\)^2
\end{align}
where the electric charge has been set to $1$.
The negative sign difference between the layers is because the exciton order parameter is composed of electron (hole) in the conduction (valence) layer.
We can perform a gauge transformation $A_{c,\mu} \rightarrow ({A_{c,\mu} - \partial_{\mu} \phi + \v Q})$ to obtain the response Lagrangian
\begin{align}
    \L_s[A_c,A_v] = \frac{\rho_G}{2} \(A_{c,\mu}-A_{v,\mu}\)^2
\end{align}
This leads to a cross term between $A_c$ and $A_v$ implying that the DC drag conductivity, where the voltage and current are measured on opposite layers, $\sigma_{cv}(\omega = 0)=\infty$. This follows from noting that the current in the valence layer $ j_{v} = -\frac{\delta \mathcal{L}_s}{\delta A_c} \sim A_c$, analogous to the London equation in superconductors, but for drag transport.

In the collinear X-DW, we have two bosonic order parameters: $\chi_{\pm} \sim e^{i\phi_{\pm}}$. Assuming the same stiffness for the two phase fields for simplicity, we obtain the effective Lagrangian, as before.
\begin{align}
    \L_c[\phi,A_c,A_v] = \frac{\rho_G}{2} \[\( \partial_{\mu} \phi_+ - \v Q - (A_{c,\mu}-A_{v,\mu})\)^2 + \( \partial_{\mu} \phi_- + \v Q - (A_{c,\mu}-A_{v,\mu})\)^2\]
\end{align}
We can eliminate the coupling between the phase and the gauge fields by performing a gauge transformation $A_{c,\mu} \rightarrow A_{c,\mu} + \partial_{\mu}(\phi_+ + \phi_-)/2$. We then obtain
\begin{align}
    \L_c[A_c,A_v] = \rho_G \(A_{c,\mu}-A_{v,\mu}\)^2
\end{align}
yielding $\sigma_{cv}(\omega=0)=\infty$ as for the spiral X-DW. 

The arguments presented above apply directly to the PDW superconductor by simply replacing $A_c-A_v$ by $A_c+A_v$ since the Cooper pair is electrically charged. One crucial consequence is that if the two layers are not separately contacted, $A_c=A_v$, the PDW superconductor indeed exhibits superflow, whereas, the X-DW phases do not, since the phase fields $\phi_{\pm}$ decouple from the gauge fields.

Though our mean-field analysis suggest the formation of collinear X-DW over spiral X-DW, it is instructive to determine whether this may be experimentally verified. The collinear and the spiral X-DW phases show the same qualitative behavior in drag transport. 
However, the presence of intra-layer charge density modulation in the case of the collinear X-DW can distinguish the two possibilities, for example by scanning tunneling microscopy. 

Finally, we discuss the response of the system to an in-plane magnetic field. The orbital coupling of the magnetic field will cause the two-dimensional momenta of the electrons in the two layers to shift in opposite directions, explicitly breaking $C_{2z}$. (For magnetic field along $\hat{x}$, the vector potential in the Landau gauge, $\v A \sim z \hat{y}$, which is opposite for the two layers located at $z \pm d/2$.)The equally susceptible wave-vectors for exciton condensation will then be $\v Q_1$ and $-\v Q_2$ for $\v Q_1 \neq \v Q_2$. In this case, we expect the collinear X-DW to get modified to a quasi-periodic ordered state where the strength of the exciton order parameter varies with a wavevector $(\v Q_1 + \v Q_2)/2$ and the phase of the order parameter varies with $(\v Q_1 - \v Q_2)/2$. 


\end{document}